\documentclass[12pt]{article} 
\usepackage{epsfig} \usepackage{amsfonts,amssymb,amsmath}
\usepackage{subfigure}
\newcommand{\al}{\ensuremath{\alpha}}

\renewcommand{\d}{\ensuremath{{\rm d}}}

 \newcommand{\be}{\begin{equation}}
\newcommand{\ee}{\end{equation}} 
\newcommand{\ba}{\begin{eqnarray}} \newcommand{\ea}{\end{eqnarray}}

\newcommand{\bra}[1]{\langle #1|}
\newcommand{\ket}[1]{|#1\rangle}

\newcommand{\lab}[1]{\label{#1}}  
\newcommand{\bib}[1]{\bibitem{#1}}

 \newcommand{\nc}{\newcommand}

\def\ie{\textit{i.e. }} 
\def\mn{{\mu\nu}} 
\def\tt{\textrm}
\def\ab{{\alpha\beta}}
\def\der{\partial}

\nc{\aaa}[3]{{\ Astron.\ Astroph.\ }{{\bf #1},({#2}){#3}}}
\nc{\advp}[3]{{Adv.\ in\ Phys.\ }{{\bf #1}({#2}){#3}}}
\nc{\apl}[3]{{Appl. Phys. Lett. }{{\bf #1}{(#2)}{#3}}}
\nc{\apj}[3]{{Astrophys.\ J.\ }{{\bf #1} {(#2)} {#3}}}
\nc{\apjl}[3]{{Astrophys.\ J.\ Lett.\ }{{\bf #1} {(#2)} {#3}}}
\nc{\app}[3]{{\ Astrop.\ Phys..\ }{{\bf #1}, {(#2)} {#3}}}
\nc{\asp}[3]{{\ Astropart.\ Phys.\ }{{\bf #1} {(#2)} {#3}}}

\nc{\cmp}[3]{{  Comm.\ Math.\ Phys.\ }{{ \bf #1} {(#2)} {#3}}}
\nc{\cqg}[3]{{  Class.\ Quant.\ Grav.\ }{{\bf #1} {(#2)} {#3}}}
\nc{\epl}[3]{{  Europhys.\ Lett.\ }{{\bf #1} {(#2)} {#3}}}
\nc{\ijmp}[3]{{ Int.\ J.\ Mod.\ Phys.\ }{{\bf #1} {(#2)} {#3}}}
\nc{\ijtp}[3]{{ Int.\ J.\ Theor.\ Phys.\ }{{\bf #1} {(#2)} {#3}}}
\nc{\jhep}[3]{{ JHEP\ }{{\bf #1} {(#2)} {#3}}} \nc{\jmp}[3]{{  J.\
Math.\ Phys.\ }{{ \bf #1} {(#2)} {#3}}} \nc{\jpa}[3]{{  J.\ Phys.\ A\
}{{\bf #1} {(#2)} {#3}}} \nc{\jpc}[3]{{  J.\ Phys.\ C\ }{{\bf #1}
{(#2)} {#3}}} \nc{\jpg}[3]{{ J.~Phys.~G:~Nucl.~Part.~Phys.~}{{\bf #1}
{(#2)} {#3}}} \nc{\jap}[3]{{ J.\ Appl.\ Phys.\ }{{\bf #1} {(#2)}
{#3}}} \nc{\jpsj}[3]{{ J.\ Phys.\ Soc.\ Japan\ }{{\bf #1} {(#2)}
{#3}}} \nc{\lmp}[3]{{ Lett.\ Math.\ Phys.\ }{{\bf #1} {(#2)} {#3}}}
\nc{\lncim}[3]{{ Lett.\ Nuov.\ Cim.\ }{{\bf #1} {(#2)} {#3}}}
\nc{\mpl}[3]{{ Mod.\ Phys.\ Lett.\ }{{\bf #1} {(#2)} {#3}}}
\nc{\nat}[3]{{  Nature \ }{{\bf #1} {(#2)} {#3}}} \nc{\ncim}[3]{{
Nuov.\ Cim.\ }{{\bf #1} {(#2)} {#3}}} \nc{\npb}[3]{{ Nucl.\ Phys.\
}{{\bf B#1} {(#2)} {#3}}} \nc{\pr}[3]{{ Phys.\ Rev.\ }{{\bf #1} {(#2)}
{#3}}} \nc{\pra}[3]{{  Phys.\ Rev.\ }{{\bf A#1} {(#2)} {#3}}}
\nc{\prb}[3]{{  Phys.\ Rev.\ }{{\bf B#1} {(#2)} {#3}}} \nc{\prc}[3]{{
Phys.\ Rev.\ }{{\bf C#1} {(#2)} {#3}}} \nc{\prd}[3]{{  Phys.\ Rev.\
}{{\bf D#1} {(#2)} {#3}}} \nc{\prl}[3]{{ Phys.\ Rev.\ Lett.\ }{{\bf
#1} {(#2)} {#3}}} \nc{\plb}[3]{{  Phys.\ Lett.\ }{{\bf B#1} {(#2)}
{#3}}} \nc{\prep}[3]{{ Phys.\ Rep.\ }{{\bf #1} {(#2)} {#3}}}
\nc{\prsl}[3]{{ Proc.\ R.\ Soc.\ London\ }{{\bf #1} {(#2)} {#3}}}
\nc{\ptp}[3]{{  Prog.\ Theor.\ Phys.\ }{{\bf #1} {(#2)} {#3}}}
\nc{\ptps}[3]{{ Prog\ Theor.\ Phys.\ suppl.\ }{{\bf #1} {(#2)} {#3}}}
\nc{\physa}[3]{{ Physica\ A\ }{{\bf #1} {(#2)} {#3}}} \nc{\physb}[3]{{
Physica\ B\ }{{\bf #1} {(#2)} {#3}}} \nc{\phys}[3]{{ Physica\ }{{\bf
#1} {(#2)} {#3}}} \nc{\rmp}[3]{{ Rev.\ Mod.\ Phys.\ }{{\bf #1} {(#2)}
{#3}}} \nc{\rpp}[3]{{ Rep.\ Prog.\ Phys.\ }{{\bf #1} {(#2)} {#3}}}
\nc{\sjnp}[3]{{ Sov.\ J.\ Nucl.\ Phys.\ }{{\bf #1} {(#2)} {#3}}}
\nc{\jetp}[3]{{ JETP\ }{{\bf #1} {(#2)} {#3}}} \nc{\yf}[3]{{ Yad.\
Fiz.\ }{{\bf #1} {(#2)} {#3}}} \nc{\zetp}[3]{{ Zh.\ Eksp.\ Teor.\
Fiz.\ }{{\bf #1} {(#2)} {#3}}} \nc{\zp}[3]{{ Z.\ Phys.\ }{{\bf #1}
{(#2)} {#3}}} \nc{\zpc}[3]{{ Z.\ Phys.\ C\ }{{\bf #1} {(#2)} {#3}}}
\nc{\ibid}[3]{{\sl ibid.\ }{{\bf #1} {#2} {#3}}}

\begin{document}

%\rightline{yymm.nnnn [hep-th]}
%\rightline{OUTP-04/18P}
\vskip 1cm

%%                      Title here
%%
\begin{center}
{\Large \bf The string wave function across a Kasner singularity }
\end{center}
\vskip 1cm \renewcommand{\thefootnote}{\fnsymbol{footnote}}

\centerline{\bf
Edmund J.~ Copeland$^1$\footnote{ed.copeland@nottingham.ac.uk}, Gustavo
Niz$^1$\footnote{gustavo.niz@nottingham.ac.uk} and Neil Turok$^2$\footnote{nturok@perimeterinstitute.ca}}
\vskip 1cm

\centerline{$^1$ \it The School of Physics and Astronomy, University of Nottingham,} \centerline{\it
University Park, Nottingham NG7 2RD, UK;  and}
\vskip .5cm

\centerline{$^2$ \it Perimeter Institute for Theoretical Physics,}
\centerline{\it 31 Caroline St N, Waterloo, Ontario N2L2Y5, Canada.}

\setcounter{footnote}{0} \renewcommand{\thefootnote}{\arabic{footnote}}

%%                      Text starts here
%%

\begin{abstract}
A collision of orbifold planes in eleven dimensions has been proposed
as an explanation of the hot 
big bang \cite{ekp, cyc, Mturok}. When the two planes are close to
each other, the winding membranes become the lightest modes of the
theory, and can be effectively described in terms of fundamental
strings in a ten dimensional background. Near the brane 
collision, the eleven-dimensional metric is an Euclidean space times a
1+1-dimensional Milne universe. However, one may expect small
perturbations to lead into a more general Kasner background. In this
paper we extend the previous classical analysis of winding membranes
to Kasner backgrounds, and using the Hamiltonian 
equations, solve for the wave function of loops with circular
symmetry. The evolution across the singularity is regular, and
explained in terms of the excitement of higher oscillation modes. We
also show there is finite particle production and unitarity is
preserved.  

\end{abstract}

\newpage

\section{Introduction}

The initial singularity problem remains an open question in Cosmology and any
model of the early Universe 
requires a resolution of this paradigm. We know general relativity
breaks down close to it but there is hope that a theory of quantum gravity can
resolve the singularity. Recently, interest has turned to the
particular case of bouncing models, where the question of how 
information propagates across the big crunch/big bang transition has
not been completely solved. For example, in
the ekpyrotic/cyclic model this transition is assumed
smooth with controlled particle production \cite{ekp,cyc}. In order
to prove or disprove such a statement, most people have used effective
field theories, constructed from string theory or 
other extensions of general relativity. There is plenty of literature
on this approach to tackle the singularity problem (see for example
\cite{gratton}), 
however, most of these effective theories break down near the
singularity, and 
perhaps one should be considering a more fundamental description beyond
general relativity, such as string/M-theory. One such approach is to
directly investigate the string equations of motion in a singular
background \cite{bulk}, with special attention being given to the Milne
universe \cite{bulkmilne, berkooz}. Some authors have used this to
argue that this particular singularity can not be resolved
\cite{berkooz}, however, the results are not conclusive and in fact
evidence from a dual description such as investigated in  \cite{craps}
seems to contradict the result. Furthermore, there is evidence that in a big
crunch/big bang transition in asymptotically Anti de Sitter spacetime,
the Conformal Field Theory (CFT) description leads to a well-defined
evolution of fields across the singularity \cite{das,hertog}. In 
\cite{Mturok}, the authors proposed a novel approach to explain the
Milne singularity using eleven dimensional membranes, which is a
natural setup for the cyclic universe, as also discussed recently in \cite{moss}. In this paper, we 
generalize this M-theory setup to more general backgrounds, corresponding to the 
homogeneous and anisotropic Kasner metrics. Such a background could
well result once we include the effects of small perturbations in the
background isotropic metric. Moreover, we also make progress in the
quantum evolution of such membranes across the singularity. 

In the M-theory model of \cite{Mturok}, the singularity is described
by two orbifold planes that collide as the eleventh dimension, which
separates them, disappears. The study focuses on the evolution of
membranes stretching from one orbifold plane to the other, but in
particular, considers winding membranes, which correspond to the
lowest Kaluza Klein modes in ten dimensions. As argued in \cite{Mturok},
the winding membranes represent  
the lightest modes and decouple from the bulk (heavy Kaluza
Klein) modes when the eleventh dimension is sufficiently small. From 
the ten dimensional point of view, these winding membranes are
described by perturbative string theory, hence they include
perturbative gravity. The classical evolution has been studied
in the case of the Milne universe \cite{niz}, and some progress has
been made to understand the quantum theory of such modes, either by
taking semiclassical approximations, such as the instanton calculations
of \cite{Mturok} or by prescriptions to linearize the classical
equations of motion, as in \cite{niz}. Furthermore, as shown in
\cite{niz2} the classical 
evolution does not acquire finite-width  $\alpha'$-corrections either far away from the
singularity or very close to it. Therefore, on either side of the
singularity there are two semiclassical regimes connected by a phase
where quantum corrections are important. In the present analysis 
the story repeats, but in this case we quantize the
action for certain membranes -- corresponding to circular strings
from the ten dimensional theory -- and show how particle production
remains finite even though higher oscillation modes are excited.

In the ekpyrotic/cyclic universe, the spacetime is well described by
the Milne universe near the orbifold collision \cite{ekp,cyc}. However, it is well
known that as the singularity is approached, any small perturbation to
the Milne universe can lead to a Kasner solution (see for example
\cite{billiards}). Therefore, we believe it is 
important to show how these winding membrane modes evolve across
Kasner metrics. The metric, $g_\mn$, for an eleven dimensional Kasner
space-time is given by 
\be\lab{kasmetric}
ds_{11}^2=-dt^2+\sum_{i=1}^{10} |\theta_0t|^{2p_i}(dx^i)^2,
\ee
where $\theta_0$ is a dimensionful positive constant, which for the eleventh
direction represents the rapidity at which the two orbifold planes
collide. The usual Kasner conditions hold in eleven dimensions:
\be\lab{kasnercond}
\sum_{i=1}^{10} p_i^2=1=\sum_{i=1}^{10} p_i.
\ee
We have chosen the singularity to be at $t=0$ and we have glued the
manifolds before and after the singularity using the absolute-value
function. 
In general, $t=0$ is a curvature singularity, and only for the
particular case of $p_i=1$ for a given coordinate, does the solution become a 
direct product of a 9d flat space-time and the Milne
universe, with $t=0$  a coordinate singularity, which simply represents the fact that we have made a bad choice of coordinates in flat
space. However, if the spatial coordinate of the Milne metric is 
compact, then the singularity is a conical singularity. This is the
case of the cyclic universe where the big crunch/big bang transition
is modeled by an orbifold collision, where the eleventh dimension is
compact with a $Z_2$ symmetry. The orbifold structure is not essential
for the present discussion, because our results only rely on a very
small compact eleventh dimension, so we will forget about this
discrete symmetry. In other words, because we are concentrating only on the Bosonic sector of the theory, where all the string models have the same field content, our results apply to either
Heterotic or IIA limit of M-theory.

When the eleventh dimension, $x^{10}$,
is small enough we can use a ten dimensional description based on the 
Kaluza-Klein reduction
\be
ds_{11}^2=e^{-2\phi/3}ds_{10}^2+e^{4\phi/3}d(x^{10})^2,
\ee
where the dilaton is given by $\phi=\frac{3}{2}p_\phi\ln |\theta_0t|$ (with
$p_{10}\equiv p_\phi$), and
the ten dimensional metric reduces to
\be\lab{ds10}
ds_{10}^2=a(t)^2 ds^2_{conf}=a(t)^2\left(-dt^2+\sum_{i=1}^{9}
|\theta_0t|^{2p_i}(dx^i)^2\right), \qquad \qquad  
a(t)=|\theta_0t|^{p_\phi/2},
\ee
where we have assumed $p_\phi>0$, implying the eleventh dimensions disappears
as $t\rightarrow 0$.
As $x^{10}\rightarrow 0$ we can think of these winding
membranes as fundamental strings on the orbifold planes feeling
the metric 
(\ref{ds10}). Alternatively, the winding membranes can be thought of 
as strings with a time-dependent tension living on the metric
$ds^2_{conf}$, as will become evident later. 
Since the string coupling is
$e^\phi$, for really small times --- close to the singularity --- the
strings hardly interact and one can take the free string action as a
good description. Therefore, we will focus our attention on the 
propagation of free strings on the ten dimensional dilaton-Kasner
background (\ref{ds10}).

The paper is organized as follows: in Section 2, we write down the
different actions for membrane excitations in eleven dimensions, and
in the following section we solve the classical equations of motion
governing only winding membranes with cylindrical symmetry. Section 3
is devoted to the quantum description of circular loops 
using a Hamiltonian approach, before we finally conclude in Section
4.

\section{Winding membranes}

Our starting point is a Polyakov type of action for a bosonic
membrane of tension $\mu_2$ in eleven dimensions
\begin{eqnarray}\lab{polac1}
S_{pol} &=&-\frac{\mu_2}{2}\int d^{3}\sigma \mathcal{L}_{pol} \\ \lab{polac}
&=&-\frac{\mu_2}{2}\int
d^{3}\sigma\sqrt{-\gamma}\left(\gamma^\ab\der_\al x^\mu \der_\beta x^\nu
g_\mn -1\right),
\end{eqnarray}
where $x^\mu$ are fields representing the position of the membrane in 
a target space with metric $g_\mn$. The worldvolume spanned by the
coordinates $\sigma^\al$ has a metric $\gamma_\ab$, and the variation
of this action with respect to $\gamma_\ab$ yields the constraint 
$\gamma_\ab=\der_\al x^\mu \der_\beta x^\nu g_\mn$, which can be
substituted back into the action to obtain the Nambu-Goto action,
\be\lab{ngact}
S_{NG}=-\mu_2\int\d^3\sigma \sqrt{-\tt{Det}(\der_\al x^\mu \der_\beta
  x^\nu g_\mn)}.
\ee
The first action is more convenient to analyze the quantum behavior
whereas the second is more useful to describe the classical evolution,
as we will show below. As explained in \cite{Mturok}, the Hamiltonian
can be constructed from the action (\ref{polac}), leading to the
constraints
\be
\mathcal{H}\equiv\pi_\mu\pi_\nu
g^\mn+\mu_2^2\mathrm{Det}(\der_{\hat{\alpha}}
x^\mu\der_{\hat{\beta}} x^\nu g_\mn)= 0, \qquad \qquad
\mathcal{P}_{\hat{\alpha}}\equiv\pi_\mu\der_{\hat{\alpha}} x^\mu= 0,
\ee
where $\pi_\mu\equiv\frac{\partial \mathcal{L}_{pol}}{\partial\dot{x}^\mu}$
are the canonical conjugate momenta to $x^\mu$, 
$\dot{x}^\mu \equiv \frac{\partial x^{\mu}}{\partial \sigma^0}$ 
 and the 
hatted indices run over the spatial dimensions of the membrane's
worldvolume. Therefore, the most general Hamiltonian is
\be\lab{hamil11d}
H=\int d^2\sigma
\left(\frac{A}{2}\mathcal{H}+A^{\hat{\alpha}}
\mathcal{P}_{\hat{\alpha}}\right),    
\ee
where the two functions $A$ and $A^{\hat{\alpha}}$ represent the
gauge freedom of the membrane's metric diffeomorphisms. We 
consider a partial gauge where the momentum is always
orthogonal to the membrane, which is equivalent to choosing
$A^{\hat{\alpha}}=0$, and will use the remaining gauge freedom to
simplify the equations of motion and obtain either classical or
quantum solutions. A winding membrane is obtained by demanding its coordinates $x^\mu$ are 
independent of one of the spatial membrane 
worldvolume coordinates (say $\sigma^2$), except for the eleventh
dimension which should be proportional to $\sigma^2$. We choose
$x^{10}=\sigma^2$ (where $\sigma^2$ runs from $0$ to
$1$), so that after integrating with respect to $\sigma^2$ in
(\ref{ngact}) we get an overall factor of $|\theta_0 t|^{p_\phi}$ in
front of the effective string action.

\section{Classical evolution}

To describe the classical evolution of a winding membrane in a Kasner
background we use the $t=\tau$ gauge in the Nambu-Goto type of action
(\ref{ngact}). Then the action reduces to 
\be
S=-\mu_2\int d\sigma d\tau 
|\theta_0 t|^{p_\phi}\sqrt{\left(1-\sum_{i=1}^9 |\theta_0 t|^{2p_i}
  (\dot{x}^i)^2\right)\sum_{i=1}^9 |\theta_0 t|^{2p_i} (\der_\sigma x^i)^2}, 
\ee
where $\tau\equiv\sigma^0$ and $\sigma\equiv\sigma^1$. For simplicity
we will assume  
$\theta_0=1$ during the calculations and then restore a general $\theta_0$
at the end. The equations of motion, in units of $\mu_2=1$, read 
\ba\lab{claseqns}
\dot{x}^i=\frac{\pi_i |t|^{-2p_i}}{\epsilon}, \qquad&& \qquad
\dot{\pi}_i=|t|^{2(p_\phi+p_i)}\der_\sigma\left(\frac{\der_\sigma
  x^i}{\epsilon}\right),\nonumber  \\
\dot{\epsilon_i}=(p_\phi+2p_i)|t|^{2(p_\phi+2p_i)} \frac{(\der_\sigma
  x^i)^2}{t\, \epsilon_i},\qquad && \qquad
\epsilon_i^2=\pi_i^2+|t|^{2(p_\phi+2p_i)} (\der_\sigma x^i)^2,
\ea
where the string energy density $\pi_0\equiv\epsilon$ is given by
\be\lab{epsilon}
\epsilon^2=\sum_{i=1}^9 |t|^{-2p_i}\epsilon_i^2.
\ee
Using the last expression we can rewrite the differential equation for
$\epsilon_i$ in the following way
\be\lab{epsilon2}
\der_t\left(\frac{\epsilon_i^2}{|t|^{2(p_\phi+2p_i)}}\right)=-
\frac{2(p_\phi+2p_i)}{t|t|^{2(p_\phi+2p_i)}}\pi_i^2, 
\ee
which will be useful later. From (\ref{claseqns}), divergent solutions
arise when at least one of the Kasner exponents in the 10d space-time
is negative enough to lead to a divergent term in the energy density
$\epsilon$ at $t=0$, as previously shown by Tolley   
\cite{tolley}. On the other hand, regular solutions across $t=0$ are
obtained if all $p_i\geq -p_\phi/2$. To avoid divergences we will assume 
\be\lab{regcond}
p_i\geq -p_\phi/2 \qquad \mathrm{for\ all\ }i.
\ee
The divergent cases
correspond to situations where, before the singularity, one spatial
dimension expands faster than the contraction of  the 10d conformal factor $a(t)$, as appreciated in (\ref{ds10}). We are more interested in situations
which are small perturbations away from the Milne universe, but still
close to it.

To construct a perturbative solution around the singularity, one can
expand the equations of motion (\ref{claseqns}) in terms of the string 
tension, as was done in \cite{niz}. Formally, one 
introduces a parameter $\lambda$ in place of the tension (i.e. $\mu \to \mu \lambda$) and
solves iteratively the equations of motion as a series in
$\lambda$. At the end, one sets $\lambda=1$. As shown in \cite{niz},
the only equation where $\lambda$ appears is $\dot{\pi}_i=\lambda
|t|^{2(p_\phi+p_i)}\der_\sigma\left(\frac{\der_\sigma 
  x^i}{\epsilon}\right)$, whose solution to zeroth order in $\lambda$ is
$\pi_i=\pi_i(0)$, where $\pi_i(0)=\pi_i(0,\sigma)$ is the loop
momentum at $t=0$. Assuming (\ref{regcond}) holds, one can integrate
equation (\ref{epsilon2}) and insert the solution of $\epsilon$ and
$\pi_i$ in the $\dot{x}_i$-equation, to obtain the zeroth order
solution  
\be\lab{zeroth}
x^i(t)\simeq x^i(0)+\int_{t_0}^t dt \frac{\pi_i(0)
  |t|^{-2p_i}}{\sqrt{\sum_j |t|^{-2p_j}\left[\pi_j^2(0)+(\partial_\sigma x^j(0))^2\,
      |t|^{2p_\phi+4p_j}\right]}}+\mathcal{O}(\lambda),
\ee
where $x^i(0)=x^i(0,\sigma)$ is the string shape at the singularity.
The last integral is finite, hence the solution is regular at $t=0$. 
In general, this integral has to be done numerically, but there are
specific string geometries or configurations where the solution can be
found analytically; this is the case of a circular string.

\subsection{Circular loops}

For the rest of the paper we will focus on the circular loop, which
as explained in \cite{niz} is the classical analogue of the dilaton
field. The simplification rests on the fact that the only dynamical
coordinate is the radius of the circle. Furthermore, to preserve the
circular symmetry, the two Kasner exponents of the plane where the
loop oscillates should be equal. Without loss of generality, we assume
the circular loop oscillates in the $xy$ plane and has a center of
mass velocity $v$ in the $z$ direction, with the ansatz
$x^i=\left(R(t)\cos(\sigma), R(t)\sin(\sigma),v\tau, 0,...,0\right)$, and
the Kasner exponents in these directions are $p\equiv p_1=p_2$ and
$p_3=p_z$. Under these assumptions the equations of motion
(\ref{claseqns}) simplify to  
\ba\lab{eqnscirc}
v=\frac{\pi_z |t|^{-2p_z+p}}{\tilde{\epsilon}}, \qquad&& \qquad
\dot{\pi}_z=0,\nonumber  \\
\dot{R}=\frac{\pi_R |t|^{-p}}{\tilde{\epsilon}}, \qquad&& \qquad
\dot{\pi}_R=-|t|^{2p_\phi+3p}\frac{R}{\tilde{\epsilon}},\nonumber  \\
\dot{\tilde{\epsilon}}=(p_\phi+2p)|t|^{2(p_\phi+2p)} \frac{R^2}{t\,
  \tilde{\epsilon}},\qquad && \qquad 
\tilde{\epsilon}^2=\pi_R^2+|t|^{2(p-p_z)} \pi_z^2+|t|^{2(p_\phi+2p)} R^2,
\ea
where $\pi_R^2=\pi_1^2+\pi_2^2$ and $\tilde{\epsilon}=t^{p}\epsilon$. 
Again, we can rewrite the differential equation for $\epsilon_i$ using
the last constraint, namely 
\be\lab{epsilon2circ}
\der_t\left(\frac{\tilde{\epsilon}^2}{|t|^{2(p_\phi+2p)}}\right)=-
\frac{2(p_\phi+2p)}{t|t|^{2(p_\phi+2p)}}(\pi_R^2+|t|^{2(p-p_z)} \pi_z^2).
\ee
Moreover, in the case of a circular loop it is not hard to find another
constraint by combining the different equations in
(\ref{eqnscirc}), given by
\be
(v^2+\dot{R}^2)|t|^{2p}+\dot{\pi}_R^2|t|^{-2(p_\phi+p)}=1.
\ee
Notice that the speed (squared) of any point in the loop is
$V^2={V^2_z+V^2_R}=(v^2+\dot{R}^2)|t|^{2p}$, which is 
unaffected by the contraction or expansion of the plane of
oscillation. After a careful analysis of the
second term in the last constraint, one can be convinced that
every point in the string reaches the speed of 
light ($V^2\rightarrow 1$) as $t\rightarrow 0$, if the inequalities 
(\ref{regcond}) hold. As a result of
this, the solutions are not time invariant and the outgoing mode is
different from the incoming one. Quantum 
mechanically this time asymmetry is the origin of particle production
and excitation of higher order oscillation modes, as we will show
later. This effect is enhanced when the center of mass momentum
$\pi_z$ vanishes. Furthermore, in the case when the bound
(\ref{regcond}) is saturated for the $p$ Kasner exponent, the 
speed of the loop will be finite and generically smaller than the
speed of light; hence there will not be a time asymmetry in the
solution and no 
particle production or higher oscillation modes will be expected 
across the singularity. This should be expected, since the effective
metric on the $xy$ plane (see equation (\ref{ds10})) neither
contracts or expands when the bound (\ref{regcond}) is saturated for $p$.

Although the set of equations (\ref{eqnscirc}) can not be solved analytically everywhere
they can be solved approximately in different regions and these results can then be compared to
the full numerical solutions. The solutions evolve similarly to those in
\cite{niz}, so we refer the reader to this previous work for
details. However, we would like to stress a few general points,
especially when more general Kasner exponents are considered and not
only the Milne case, as it was done in \cite{niz}.

Far away from the singularity, the string does not feel the
contraction or expansion of the universe, and therefore, it
oscillates as if it lived in the $ds_{conf}^2$ metric of equation
(\ref{ds10}), which reduces to flat spacetime for the Milne
universe. Following the
notation of \cite{niz}, at a time $t_0$  
the winding membrane in eleven dimensions can be effectively described
by perturbative string theory in ten dimensions.  Furthermore, the
string coupling $e^\phi=|\theta_0 t|^{3p_\phi/2}$ tends to zero as the
singularity is approached, hence the free string action
becomes more accurate closer the orbifold collision. By definition, $t_0$
corresponds to the time where the string coupling
hits unity, namely 
\be\lab{t0}
t_0=\theta_0^{-1}.
\ee
The string tension is therefore $\mu_1=|\theta_0 t_0|^{p_\phi/2}=1$
and its length is $l_s\sim\mu_1^{-1}=1$. In terms of a quantum
analysis we would expect this regime to be well described by a
semiclassical solution, which can be obtained using the WKB 
approximation. However, classically we start with a string
configuration in the metric $ds_{conf}^2$ at time $t=-t_0$ and evolve it
towards the singularity. The solution crosses the singularity
and after it has reached a large enough positive time (comparable with
$t=+t_0$), we can trust the description of strings living on the 
metric $ds_{conf}^2$ again. We aim to compare both states, the ingoing
and the outgoing states at $t=\pm t_0$ respectively, to determine
whether there was particle production or excitation of higher
vibrational states. Quantum mechanically, it
corresponds to calculating a mini S-matrix, defined by the evolution
of the quantum ingoing states at $t=-t_0$ to the outgoing ones at
$t=+t_0$. 

Starting from a large negative time in this adiabatic
vacuum ($t=-t_0$), the string evolves into the singularity increasing
its size, according to the contraction of the conformal factor of the
universe. By a simple rescaling of the coordinates it is possible to 
show that the loop scales as a power of $1/a(t)\sim t^{-p_\phi/2}$. Once
the size of the 
loop is comparable to the averaged conformal Hubble radius ($\sim
1/|t|$), the ``stringy'' quantum corrections become really important and the
evolution can no longer be described by a 
semiclassical analysis. Remember the string coupling goes to zero
as $t\rightarrow 0$, hence only the $\alpha'$-corrections become
important as the solution approaches the singularity. As shown in \cite{niz2},
these $\alpha'$-corrections modify the 
semiclassical evolution, but in a finite way. Finally, close to the 
singularity, there is another semiclassical phase in which the modes
freeze, stop oscillating and cross the singularity. In detail, the
string ``breaks'' into string bits which evolve independently of each
other. In other words, the spatial gradients that tie the string
together become negligible and the evolution only depends on
time. This phenomenon is a consequence of the ultra-local behavior
that one expects near a cosmological singularity (see for example
\cite{billiards}). At
$t=0$, however, the string receives an energy kick, because it has to
travel at the speed of light, which either increases or decreases the
amplitude of the outgoing mode, leading to classical gain or loss of 
energy. Quantum mechanically this translates into particle production,
as we will discuss in the next Section. 

Different Kasner exponents can dramatically change the outcome because
they play an important role close to the singularity. In order to
simplify the discussion, we will only consider the $\pi_z=0$ case, but
the analysis can be easily generalized for a non-zero center of mass
momentum. Therefore, from the equations of motion  
(\ref{eqnscirc}) one may consider the exponents $p$ and $p_\phi$ as
two free parameters of the model, however, the 11d 
constraints (\ref{kasnercond}) relate these parameters, together with
the exponents of the orthogonal directions. To simplify the argument,
one may think of the orthogonal exponents as being split into two
sets, the first being  
all the same ($m$
of them) and the other set being 
zero ($6-m$ of them), \ie $p_i=\bar{p}$ ($i=3,...,m+3$) and $p_j=0$
($j=m+4,...,9$). Then, after solving for $p_\phi$ in terms of $p$ and $m$
using the constraints (\ref{kasnercond}), one obtains
\be \lab{pphisol}
p_\phi=[(1-2p)\pm \sqrt{m(m+4p-2(m+3)p^2)}]/(m+1).
\ee 
To get real
exponents, $p$ should lie between $(1-\sqrt{1+m(m+3)/2})/(m+3)\leq
p\leq (1+\sqrt{1+m(m+3)/2})/(m+3)$. There is therefore a wider range of
allowed values for $p$ as $m$ increases, but also divergent solutions
are more likely to arise since $p_\phi$ may be smaller than $-2p$,
and especially for the negative branch of the $p_\phi$ solution above
(see Figure \ref{kasnerplots}). 

\begin{figure}[t!]
\begin{center}
\resizebox*{3.2in}{2in}
{\includegraphics{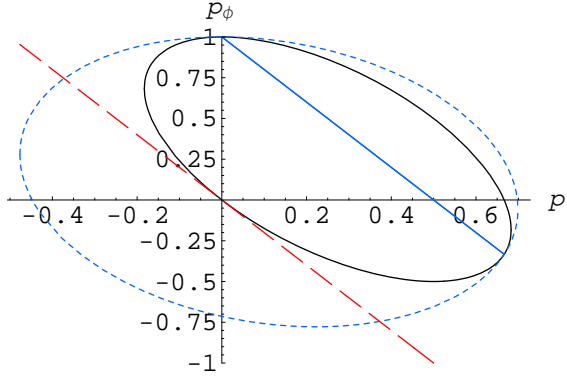}}
\caption{Kasner exponent $p_\phi$ as a function of the Kasner
  exponents in the $xy$ plane ($p$) and the $m$ equal-valued
  Kasner exponents in the orthogonal directions. The blue solid straight 
  line is for $m=0$, the solid line ellipsis is for $m=1$ and the
  blue dotted one for $m=6$. Divergent solutions which cannot be followed
  across the singularity correspond to values of $p_\phi$ 
  below the red dashed line which only exist for $m>1$.}
\lab{kasnerplots}
\end{center}
\end{figure}

\begin{figure}[t!]
\begin{center}
\resizebox*{4.8in}{3in}
{\includegraphics{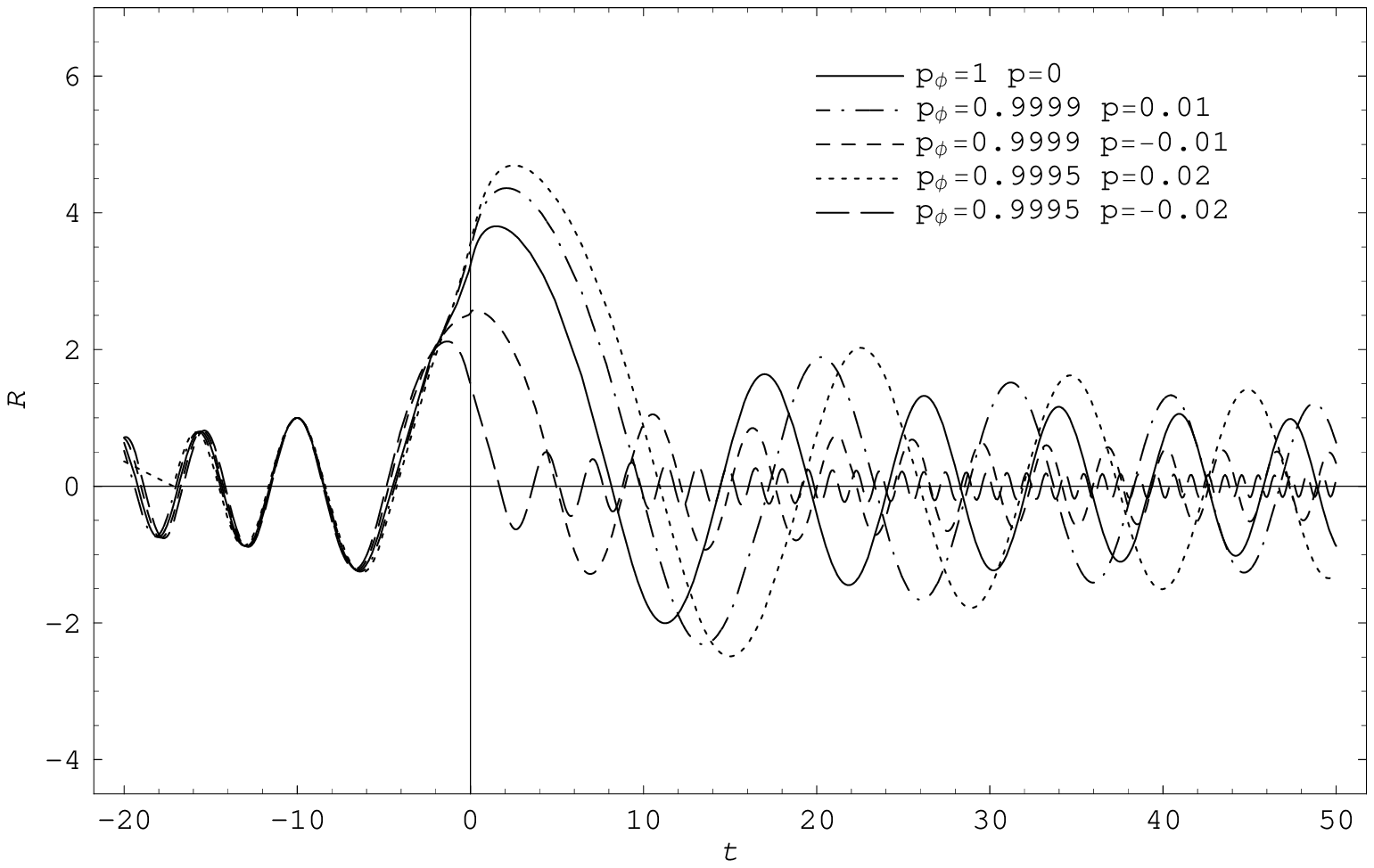}}
\resizebox*{4.8in}{3in}
{\includegraphics{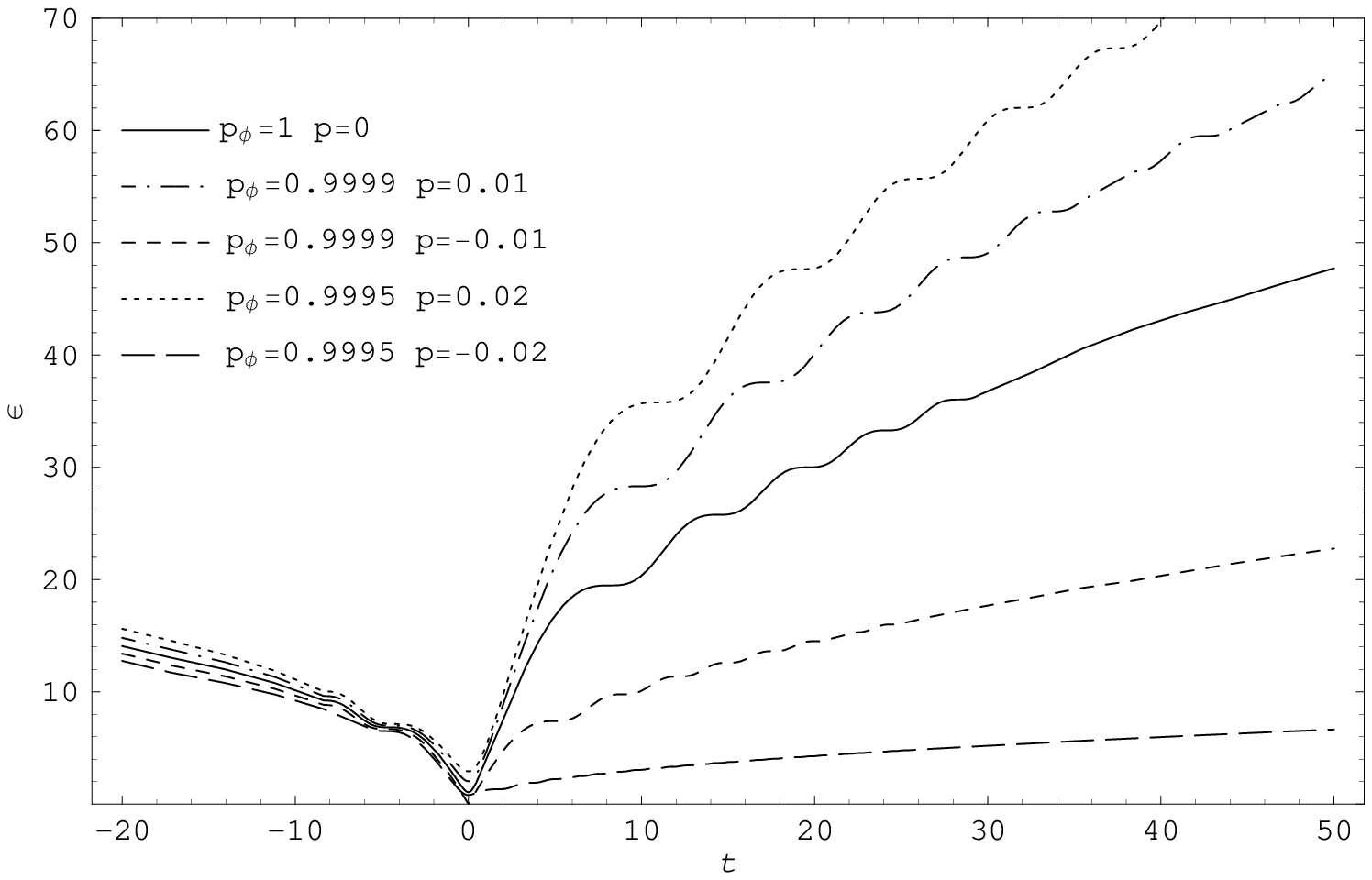}}
\caption{The radial coordinate $R$ and the energy density $\tilde{\epsilon}$
  evolving in time $t$ for different Kasner exponents around the
  Milne solution ($p_\phi=1$ and $p=0$). The evolution is regular
  across $t=0$, when the energy density takes its lowest value. The 
  solutions are not time reversal and strongly depend on the Kasner
  exponents. We have assumed $\pi_z=0$, and the initial conditions are
  $R(-t_0)=1$ and $\dot{R}(-t_0)=0$, with $t_0=20$.}
\lab{plots}
\end{center}
\end{figure}

\begin{figure}[t!]
\begin{center}
\resizebox*{4.8in}{3in}
{\includegraphics{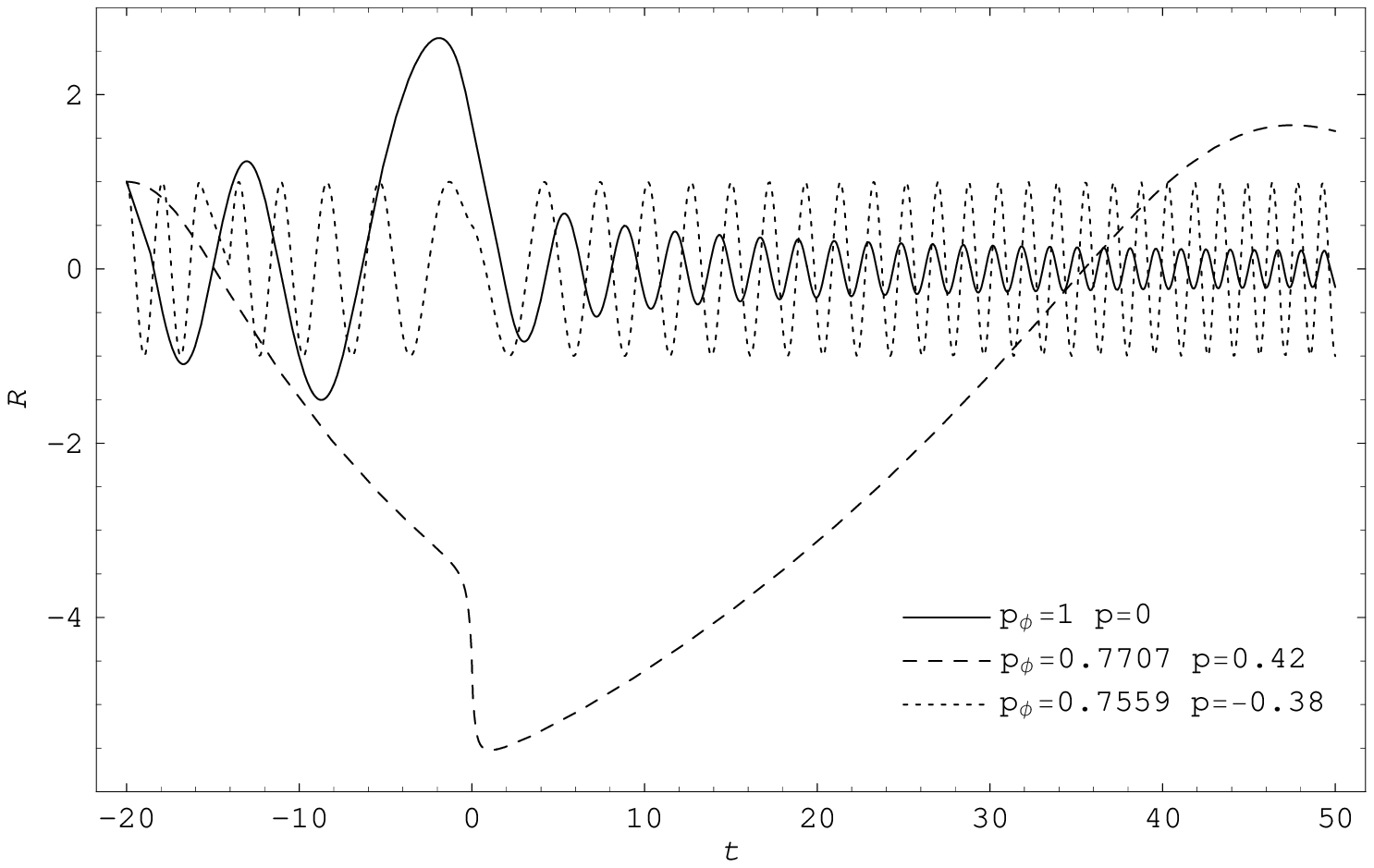}}
\caption{The radial coordinate $R$ evolving in time $t$ for different
  Kasner exponents far away from the Milne solution ($p_\phi=1$ and
  $p=0$). For positive $p$ the kick around the singularity is stronger
  and the outgoing solution is drastically modified. On the other hand,
  negative values for $p$, close to saturating the bound
  (\ref{regcond}), lead to relatively little modification of the outgoing solution, resulting
  in less particle production, as will be explained in the 
  quantum model. We have assumed $\pi_z=0$, and the initial
  conditions are $R(-t_0)=1$ and $\dot{R}(-t_0)=0$, with $t_0=20$.} 
\lab{moreplots}
\end{center}
\end{figure}

To have a feeling of how sensitive the solutions are to the Kasner
exponents let us vary $p$ around the Milne solution ($p=0$ and
$p_\phi=1$). Figure (\ref{plots}) shows a slight variation around the
Milne universe, and even though the general behavior remains
similar, the amplitude and periodicity of the outgoing modes strongly
depends on the precise value of the Kasner exponents. 
Figure (\ref{moreplots}) shows a more dramatic change
near the singularity when the Kasner exponents are taken to be far
away from the Milne universe. If $p\sim-p_\phi/2$, the energy density 
has a very mild dependence on $|t|$, so the modes do not feel the
contraction or expansion of the universe. Therefore, there is no 
classical energy production, and as we will see later, there is also a
mild particle production in the quantum theory, which consistently
tends to zero for $p=-p_\phi/2$. The opposite case, when $p$ is
positive and relatively large, the string feels effectively a larger
contraction/expansion of the scale factor, which produces a big effect
around $t=0$, and thus a greater quantum production of particles.

As mentioned before, we can describe the evolution of these strings as an
expansion in the string tension around $t=0$. Formally, we
introduce a parameter $\lambda$ where the string 
tension is and then truncate the expressions to the desired order in
$\lambda$.  Finally we set $\lambda=1$. This was done above explicitly
for a general string configuration to zeroth order in the string
coupling, and in the case of circular loop (with $\pi_z=0$) the
integral (\ref{zeroth}) can be solved analytically to give:
\be\lab{sol0}
R=R_0+t\,\frac{\mathrm{Sign}(\pi_R^0)}{(1-p)|t|^{p}}
\ {_2\mathrm{F}_1}\!\!\left[\frac{1-p}{2(p_\phi+2p)}, \frac{1}{2};
1+\frac{1-p}{2(p_\phi+2p)};
\frac{R_0^2}{\pi_R^0}|t|^{2(p_\phi+2p)}\right]+\mathcal{O}(\lambda), 
\ee
where $R_0$ and $\pi_R^0$ are the values of $R$ and $\pi_R$ at $t=0$,
and $_2$F$_1$ is the Gauss hypergeometric function, defined as the
series
\be
_2\mathrm{F}_1[a,b;c;z]=\sum_{n=0}^\infty \frac{(a)_n(b)_n}{(c)_n n!}z^n,
\ee
with $(w)_n\equiv w(w+1)...(w+n-1)$ for any complex number $w$. The
hypergeometric function is well behaved for all values of the Kasner
exponents that satisfy the bound (\ref{regcond}). Solution (\ref{sol0})
reduces to that found in \cite{niz} for $p_\phi=1$ (and $p=0$).

\section{Quantum description}\lab{quantumsection}

%% Studying the quantum picture of one a free string
%% in a Kasner background is very complicated, but we can make
%% progress by considering only circular loops and the same Kasner
%% exponent for the two directions of the plane where the string
%% oscillates. 
%% As argued in \cite{niz} a circular loops should correspond to the
%% dilaton field, hence we are describing a free scalar field across the
%% singularity. We hope this simplified model can enlighten the path to
%% follow to quantize a more general string configuration. Therefore, we
%% will assume
%% $x^\mu=R(\sigma^0)(f(\sigma^1),\cos(\sigma^1),\sin(\sigma^1),0,...,0)$,  
%% where without loss of generality we let the string to oscillate on the
%% $xy$ plane. Notice that the circular symmetry forces
%% $\der_{\sigma^0}x^0=0$, which implies $f(\sigma^1)$ is a constant and
%% we can use $\sigma^0=\tau$ to parametrize time.
%% Therefore, for a circular winding membrane on a Kasner background with
%% $p_x=p_y=p$, the Hamiltonian constraints simply to
%% \be
%% \mathcal{H}\equiv \pi_\mu\pi_\nu g^\mn+\mu_2^2\mathrm{Det}(\der_i
%% x^\mu\der_j x^\nu)\approx 0, \qquad \qquad
%% \mathcal{P}_i\equiv\pi_\mu\der_i x^\mu\approx 0,
%% \ee

We now turn our attention to the more complicated problem of
quantization. If one tries to na\"{\i}vely quantize the classical
equations (\ref{claseqns}), all sorts of problems arise, because of the
 square root present in the action. However, one can take a different approach,
by considering  the Polyakov type action and using the Wheeler-de Witt
formalism, namely 
\be\lab{wheeler-de-witt}
\hat{H}\Psi=0,
\ee
where $\Psi$ is the wave function of the string. However, it is hard
to proceed from here, because the Hamiltonian includes a term
proportional to $\der_\sigma x^0\equiv \der_\sigma t$, so it is
difficult to synchronize the different points on the string and to
talk about a common time for the string. Fortunately, if we restrict
ourselves to circular strings where the only degree of freedom 
is the radius of the loop, $R$, then the Hamiltonian simplifies
enough for the problem to be tackled. In order to preserve the
circular symmetry in time, we also need the Kasner exponents of the
plane where the loop oscillates to be equal, as we assumed in the
previous section. Moreover, a circular loop can have an initial
  center of mass momentum perpendicular to the plane of
oscillation. We take the circular winding membrane ansatz
$x^\mu=\left(t(\tau,\sigma^1), R(\tau)\cos(\sigma^1),
R(\tau)\sin(\sigma^1),v\tau,0,...,0, \sigma^2\right)$,   
where without loss of generality we allow the
string to oscillate in the $xy$ plane, with $p_1=p_2=p$, and a
constant center of mass velocity $v$ in the $z$ direction with
Kasner exponent $p_3=p_z$. Notice that the circular symmetry forces 
$\der_{\sigma}t=0$, which implies $t$ is a function of
$\tau$ only. Then the Hamiltonian (\ref{hamil11d}), reduces to 
\be
H=\int d^2\sigma \frac{A}{2}\left[-(\pi_0)^2+ |\theta_0t|^{-2p} \pi_R^2+|\theta_0t|^{-2p_z} \pi_z^2
  +|\theta_0t|^{2(p_\phi+p)}R^2\right], 
\ee
where, as mentioned before,
we have chosen the partial gauge $A^{\hat{\alpha}}=0$. To simplify
this expression we can perform a 
canonical transformation to a new time coordinate given by $\tilde{t}\sim
t^{1+p}$, up to a constant factor. Finally, because the Hamiltonian 
density does not depend on the spatial worldvolume variables, we can
choose the extra gauge freedom to fix $A=2 t^{2p}/(\int
d^2\sigma)$, simplifying the Hamiltonian to
\be\lab{hamcom}
H=-(\tilde{\pi}_0)^2+
\pi_R^2+|\theta_0\tilde{t}|^{2(\tilde{p}-\tilde{p}_z)}\pi_z^2+|\theta_0\tilde{t}|^{2(\tilde{p}_\phi+2\tilde{p})}R^2,  
\ee
where $\tilde{\pi}_0={t^p}\pi_0$ is the canonical momentum of
$\tilde{t}$, and the new Kasner exponents are
\be\lab{newkasner}
\tilde{p}_\phi=p_\phi/(1+p),\qquad \qquad \tilde{p}=p/(1+p), 
\ee
where we assume $p \neq -1$ (which is consistent with (\ref{regcond})). In fact, it is possible to rescale
$\tilde{t}$ and $R$ (preserving the Poisson brackets) in such a way
  that $\theta_0$ only appears in the $\pi_z$ term. Then, one can
  redefine the center of mass momentum to absorb the information of
  both, the orbifold rapidity $\theta_0$ and the loop's center of mass momentum $\pi_z$,
  into a single {\it combined} parameter, $\pi_o$, in the following way  
\be\lab{pi_c}
{\pi}_c=\theta_0^{(2\Delta p +s)/(s+2)}\pi_z,
\ee 
where
\be\lab{sdef}
s\equiv\tilde{p}_\phi+2\tilde{p},\qquad \Delta p\equiv\tilde{p}-\tilde{p}_z.
\ee
Notice that making $\theta_0$ larger increases $\pi_c$
if $\Delta p>-s/2$, which is true within some open set of the allowed
parameter space of the Kasner exponents. Evidently, this open set includes the Milne
universe ($\tilde{p}=\tilde{p}_z=0$ and $\tilde{p}_\phi=1$). So in
order to understand the physics of the cyclic model 
with respect to the orbifold rapidity, in what follows, we will
consider Kasner combinations which obey $2\Delta p>-s/2$.

However, solutions of the wave function using the Hamiltonian
(\ref{hamcom}) are messy when the term with $\pi_c$ is
present, thus we will focus our attention to the limit
$\pi_c\rightarrow 0$ first and then present the results for a
non-vanishing $\pi_c$. One can neglect $\pi_c$ in two different
limits: either when the orbifold 
rapidity is very small compared to the loop's momentum, or when the center
of mass velocity is small enough on its own\footnote{Note that
  $\theta_0<1$ is needed to start with loops inside the Hubble horizon
  during the contracting phase, as shown in \cite{niz}, so that in
  this case $\pi_z\simeq 0$ implies $\pi_c\simeq 0$ (where we have also
  assumed $2\Delta p>-s/2$).}. 

\subsection{Zero center of mass momentum}

When dropping the center of mass momentum term, some observables, such as the amount of 
particle production, do not depend explicitly on the orbifold
rapidity, which is given by $\theta_0$. This statement may
sound like a contradiction because we would expect more particles
being produced for larger collision rapidities, but if one looks in more
detail, the result is consistent with QFT in curved space. The basic
idea of particle production is to measure the ``difference'' between two
vacuum states, which in our case correspond to one in the far past, before the
singularity, and another in the far future. However, a change in
$\theta_0$ not only changes the orbifold rapidity but also the two
vacua we are comparing, and for the case of $\pi_c\rightarrow 0$ the
changes are such that their effects cancel each other.

Therefore, knowing that we can  absorb
$\theta_0$ by a constant rescaling of the variables in the case of
negligible $\pi_c$, we will set $\theta_0=1$ and then restore it in
the final expressions. Thus, for the present discussion we consider the
simpler Hamiltonian 
\be\lab{newham}
H=-(\tilde{\pi}_0)^2+
\pi_R^2+|\tilde{t}|^{2s}R^2,
\ee
%Note that $s\geq 0$ for regular classical solutions across $t=0$,
%which follows from (\ref{regcond}).

%% For a non-vanishing center of mass
%% momentum $P_z$ at $t=-t_0$, rescaling $\theta_0$ away is not possible, hence
%% physical observables would definitely depend on the value of the orbifold
%% rapidity, $\theta_0$. However, as explained in Appendix A, the Hamiltonian
%% (\ref{newham}) is a good approximation for very small $\theta_0$,
%% compared to the loop's momentum $P_z$, at $t=-t_0$. Therefore, we will
%% continue studying the zero center of mass momentum case, having in
%% mind that it is a good approximation for the more general case,
%% provided $\theta/P_z\ll 1$.  The opposite limit will be studied in
%% Appendix A.
A canonical quantization ($\tilde{\pi}_0\rightarrow i\der_{\tilde{t}}$ and
$\pi_R\rightarrow i\der_R$) implies that the Wheeler-de Witt equation
(\ref{wheeler-de-witt}) reduces to 
\be\lab{eqn}
[\partial_t^2- \partial_R^2 +|t|^{2s}R^2]\Psi(t,R)=0,
\ee
where we have dropped the tilde over the time variable for simplicity,
but the reader should remember that we are actually still referring to
$\tilde{t}$ defined earlier. Eqn.~(\ref{eqn}) corresponds to the
Klein-Gordon equation with the 
potential of a harmonic oscillator with a time-dependent frequency
$\omega(t)\sim |t|^{s}$. There  is a parallel line of thought to
understand this result: as was mentioned earlier, an alternative 
description of a string evolving across the metric (\ref{ds10}) is to
think of a string oscillating in the space-time $ds_{conf}^2$ with a
tension that varies with respect to time (\ie $\mu_1=\mu_2|\theta_0
t|^{p_\phi}$). From this point of view, a natural wave equation for 
the string is that of a harmonic oscillator in the metric
$ds_{conf}^2$ with a time-dependent mass, which after a change of
variables can be recast into a time-dependent frequency problem (see
for example \cite{pedrosa}). Furthermore, when
solving string equations in singular backgrounds, we generally find
harmonic oscillator equations with a time-dependent frequency
(see for example \cite{craps2009}). 

\subsubsection{Asymptotic solution}

The first step to construct a solution to 
equation (\ref{eqn}) is to understand the asymptotic behavior, where
the adiabatic or WKB approximation provides a good approximation to the true solution. 
Na\"{\i}vely we would
expect this to be in the adiabatic regime when $|t|$ is large enough,
however, as we will see later, the story is a bit more complicated and
the adiabatic regime holds only for large $T$ time ($T\sim
t^{(2+s)/2}$) \footnote{Remember the effective (2+1)-metric where
  the string oscillates is given by equation (\ref{ds10}), namely
  \mbox{ $ds^2_{3d}=|t|^{p_\phi}(-dt^2+|t|^{2p}dx^2+|t|^{2p}dy^2)=\tilde{|t|}^{(\tilde{p}_\phi+2\tilde{p})}(-d\tilde{t}^2+dx^2+dy^2)$}, 
  where $\tilde{t}\sim t^{p+1}$ and $\tilde{p}_\phi$ and $\tilde{p}$
  given by (\ref{newkasner}). 
  Therefore, the ``10-dimensional'' time $T$ is simply given by $T=\int
  dT\sim\int \tilde{t}^{s/2}d\tilde{t}$, with $s=\tilde{p}_\phi+2\tilde{p}$. }. 
Nevertheless, we can still get a feeling of the asymptotic    
solution if we set the frequency to be constant, namely  
%% In terms of the
%% harmonic oscillator with a time-dependent frequency $\omega$, we  
%% are considering a regime where the harmonic oscillator frequency 
%% remains almost constant (\ie $\dot{\omega}/\omega^2<<1$) and the
%% space-time curvature is negligible (\ie $ds_{conf}^2\sim
%% ds_{flat}^2$), so that the equation (\ref{eqn}) can be approximated by 
\be\lab{hamitoneqn}  
[\partial_t^2- \partial_R^2 +t_0^{2s}R^2]\Psi(t,R)=0,
\ee
where $t_0>>1$. The wave function that solves this simplified equation
is related to the usual harmonic oscillator form, and is given by
\be\lab{asym-wave-func}
\Psi(t,R)=\sum_{n=0}^{\infty} H_n(t_0^{s/2}R) \exp(-t_0^{s}
  R^2/2)\left[A\exp(iE_n t)+B\exp(-iE_n t)\right],
\ee
where $E_n^2\equiv (2n+1)t_0^{s}=(2n+1)\omega(t)$ and $H_n(x)$ is
the Hermite polynomial of degree $n$. Notice that the energy levels
$E_n$ 
are actually the square root of the usual harmonic oscillator
levels. To understand this, one can think of the
classical string analysis previously done, where in fact, the
Hamiltonian $\tilde{\epsilon}$ (see equation (\ref{eqnscirc})) is the square
root of the harmonic oscillator for constant time.

The asymptotic solution (\ref{asym-wave-func}), provides a hint as to
the best ansatz we can adopt to obtain the general solution to
($\ref{eqn}$). Using the harmonic oscillator as a 
basis and replacing $t_0$ by $t$, we consider an ansatz of the form: 
\be\lab{harmonicbasis}
\Psi(t,R)=\sum_{n=0}^{\infty} A_n(t) H_n(|t|^{s/2}R) \exp(-|t|^{s}
  R^2/2),
\ee
where the $A_n$'s are considered to be functions of time and determine the evolution
of the harmonic oscillator states from the incoming vacuum at negative
times to the outgoing modes far after the singularity. We can then use
the orthogonality of the Hermite polynomials 
\be\lab{ortho}
\int H_n(x)H_m(x) e^{-x^2/2}dx=\sqrt{\pi}(2^n n!)\delta_{mn}
\ee
and the relations $H'_n(x)=2n H_{n-1}(x)$ and
$H_{n+1}(x)=2xH_n(x)-2nH_{n-1}(x)$ to decompose 
equation ($\ref{eqn}$) into an infinite system of coupled ODEs for the
$A_n$'s, which are given by the following recursive equation
\ba\lab{A-eqn}
0&=&\ddot{A}_n-\frac{s}{2t}\dot{A}_n+
\left[4s-s^2(1+2n+2n^2)+16(1+2n)|t|^{2+s}\right]\frac{A_n}{16 |t|^2} 
\\ && +(n+1)(n+2)\frac{s}{t}\dot{A}_{n+2}
-(2+s)(n+1)(n+2)\frac{s}{4|t|^2}A_{n+2}
-\frac{s}{4t}\dot{A}_{n-2}
\nonumber\\ &&+(2+s)\frac{s}{16|t|^2}A_{n-2}+ \nonumber  
(n+1)(n+2)(n+3)(n+4)\frac{s^2}{4|t|^2}A_{n+4}  +\frac{s^2}{64|t|^2}A_{n-4}.
\ea
One should note that these equations are not
regular at $t=0$, so one cannot find solutions which interpolate
between negative times and positives ones. This implies nothing else than
using the harmonic oscillator basis is not a good approximation around
the singularity. However, as we will see in the next section the
evolution across the singularity is simpler than one could possibly have
expected. 

Consider the zeroth mode $A_0$ equation,
\be\lab{A0eqn0}
\ddot{A}_0-\frac{s}{2t}\dot{A}_0+\left(4s-s^2+16|t|^{2+s}\right)A_0
+\frac{2s}{t}\dot{A}_2-\frac{s(2+s)}{2t^2}A_2
+\frac{6s^2}{t^2}A_4=0.  
\ee
If we assume the higher modes are negligible, $A_n=0$ for $n>0$ (which
should be the case if the incoming state is the vacuum and will be
justified later), then Eqn.~(\ref{A0eqn0}) can be solved in closed form
for either positive or negative times, resulting in
\be\lab{A0sol}
A_0=|t|^{(2+s)/4}(\theta_0)^{s/4}\left[c_1K^{(1)}_{l}\left(2k \theta_0^{s/2}
    |t|^{(2+s)/2}\right)+c_2K^{(2)}_{l}\left(2k \theta_0^{s/2}
    |t|^{(2+s)/2}\right)\right],
\ee
where $K^{(1)}_l(x)$ and $K^{(2)}_l(x)$ are the Hankel functions of the
first and second kind respectively, $c_1$ and
$c_2$ are integration constants, $k\equiv 1/(2+s)$, $l\equiv k
\sqrt{1+s^2/2}$, and we have restored $\theta_0$ in the
expression. Using the asymptotic expansion of the Hankel functions
$K_\alpha^{(1,2)}(x)=\sqrt{\pi/(2x)}\exp\left(\pm i(x-(2\alpha\pm 1)\pi/4)\right)$, we obtain an asymptotic solution for the
ground state given in terms of positive and negative frequency
components, namely
\be\lab{vacuumsol}
\Psi(t,R)= exp(-|\theta_0t|^{s}
  R^2/2)\left[C_1\exp(iE_0 |t|)+C_2\exp(-iE_0 |t|)\right],
\ee
where $C_1=\sqrt{\frac{\pi}{2}}c_1\exp(-i(2\alpha+ 1)\pi/4)$,
$C_2=\sqrt{\frac{\pi}{2}}c_2\exp(i(2\alpha-1)\pi/4)$ and now the
frequency is time-dependent and given by
$E_0(t)=\sqrt{\omega_0(t)}=\frac{2}{2+s}|\theta_0 t|^{s/2}$. 

To justify dropping the higher order modes ($A_n$ with $n>0$) in
equation (\ref{A0eqn0}) for large times one can re-write the recursive
differential equation in terms of the $T$ time (
$T=\frac{2}{2+s}\theta^{s/2}t^{(2+s)/2}$) and 
then see which are the leading terms
for large $T$. Equation (\ref{A-eqn}), in $T$-time and with 
$\theta_0=1$, reads 
\be
0=\frac{d^2{A}_n}{dT^2}+(1+2n)A_n+\mathcal{O}\left(\frac{1}{T}\right),
\ee
and has a  large $T$ solution, 
\be\lab{En}
A_n(t)=\exp(i\sqrt{2n+1}T)=\exp(iE_nt), \qquad \qquad
E_n(t)=\frac{2\sqrt{2n+1}}{2+s}|t|^{s/2}.
\ee
The picture
is then the following: asymptotically (large $T$), all the harmonic oscillator modes
decouple from each other and as one approaches the singularity they
start interacting. If one starts with the ground state as the
incoming vacuum, then higher order modes become excited in order to
resolve the singularity, and one ends up with a tower of states as the
outgoing state in the far future, when again the interaction of
harmonic oscillator modes stops and we can use this basis to describe
the resulting state. Furthermore, since the $A_n$-equation
(\ref{A-eqn}) is a ``double-step'' recursive equation, if one starts
with the ground state in the far past, then only even $n$-levels will
be excited across the singularity.

\subsubsection{Solution near the singularity}

If we consider the vacuum positive energy mode (\ref{vacuumsol}) as the
incoming state for $t\rightarrow-t_0$ (with $t_0>>1$), then we can 
evolve numerically the solution to the full equation (\ref{eqn}). The
numerical solution behaves regularly everywhere, particularly at
$t=0$. In order to picture the evolution of the wave function across
the singularity, we can use the integrated wave equation
\be\lab{intewave}
\Psi(t)=\int_{-\infty}^{+\infty}\Psi(t,R)dR.
\ee
The numerical solution of such an integrated function is shown in Figure
\ref{wavefn}, and notice in particular that around $t=0$ the solution
is simply given by a straight line, as we shall shortly describe. This
result provides a simple explanation of how the modes evolve in terms of the
absolute value function.
The fact that the wave function (\ref{intewave}) is well approximated
by $\Psi(t)=at+b$ around $t=0$, can be understood as follows: assume
$\Psi$ behaves like $\Psi(t,R)=e^{-|t|^{s}R^2/2}f(t)$ near
the singularity\footnote{If one plots the wave function near $t=0$ as a function
  of $R$, one finds a good fit using the Gaussian profile
  $exp(-|t|R^2/2)$, which justifies our assumption.}. After substituting  
it into the Hamiltonian equation (\ref{wheeler-de-witt}), and assuming a
simple polynomial function for $f(t)$ of the form $f(t)=at^m+c$, then
the only solution with a well-defined limit as $t\rightarrow 0$ has
$c=0$ and $m=s/2,\ (2+s)/2$. Therefore, the 
solution near $t=0$, which is valid on both sides of the singularity,
is 
\be\lab{psi-straight}
\Psi(t,R)=e^{-|t|^{s}R^2/2}(at+b)|t|^{s/2}/\sqrt{2\pi},
\ee
which after integrating over $R$ implies $\Psi(t)=at+b$.  If we
think of the harmonic oscillator basis, the functions $A_n$ should all
behave as $(a_nt+b_n)|t|^{s/2}$ for small times on either side of the
singularity. However, it doesn't mean we can simply solve for the individual 
coefficients $a_n$
and $b_n$, because from (\ref{harmonicbasis}) and (\ref{intewave}) we require knowledge of 
all the modes that cross the singularity, thus no individual mode can be evolved across using this description
for the integrated wave function around the singularity. Fortunately,  from the
numerical solution, we can always calculate the $A_n$ coefficients
using the orthogonally property (\ref{ortho}) of the Hermite
polynomials. Therefore, we can show the agreement between the numerical
solution and the expected behavior ($(a_nt+b_n)|t|^{s/2}$). For example,
Figure \ref{An2} shows the case of the zeroth mode $A_0$ before the
singularity.

\begin{figure}[t!]
\begin{center}
\resizebox*{4.8in}{3in}
{\includegraphics{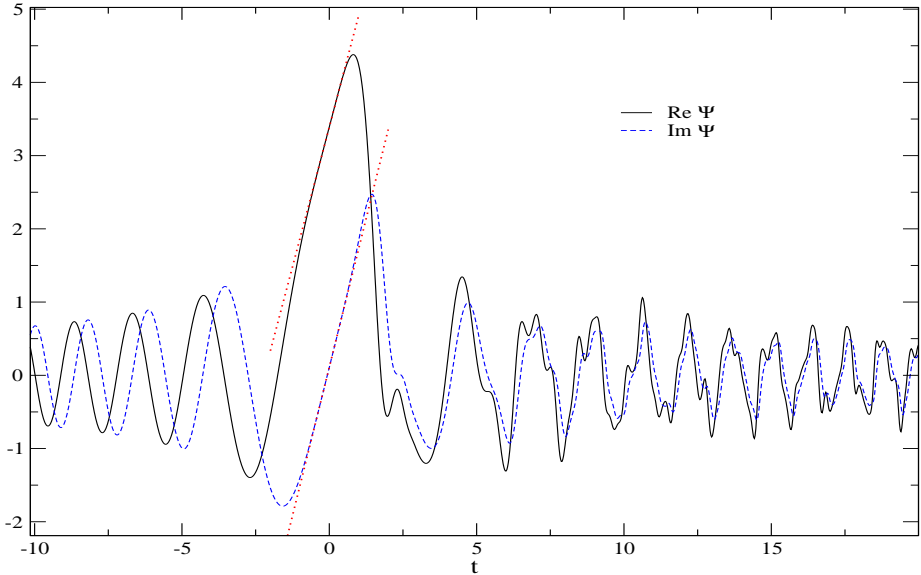}}
\caption{Evolution of the integrated wave function $\Psi(t)$ in time
  for the Kasner exponent combination $p_\phi=0.793521$ and
  $p=0.4$. The positive frequency mode in (\ref{vacuumsol}), with
  $C_1=1$ and $C_2=0$, at $t=-20$ was chosen. Different
  initial times do not change the behavior but the overall scale.
  Around $t=0$ the wave function evolution follows a straight line,
  and after the singularity, higher order modes contribute to the
  overall wave function, leading to a non-periodic and complicated
  structure.}   
\lab{wavefn}
\end{center}
\end{figure}

On the other hand, after $t=0$ the $R$-integrated wave equation
$\Psi(t)$ no longer looks periodic (see Figure \ref{wavefn}). However,
all of this complicated structure can be decomposed as a tower of higher
harmonic oscillator modes, which were excited during the transition
through $t=0$ and converge as we increase the mode number (see Figure
\ref{An}). The choice of Kasner exponents has an effect on
the tower of excited states, as will be shown later when calculating
the particle production.

Furthermore, if we follow the whole evolution of the vacuum state, the
adiabatic regime is well approximated by the analytical 
solution (\ref{A0sol}), even as we approach the singularity. However,
such an agreement inevitably breaks down very close to the singularity, as the expansion
in terms of uncoupled harmonic oscillator states breaks down.  That is, all
higher modes begin to get excited in order to solve through the singularity, hence,
we can not neglect the $A_2$ or $A_4$ terms in the $A_0$
equation near $t=0$.  However, notice that these higher order terms never dominate, as can be seen in Figure \ref{An}. As the adiabatic approach fails,
the above mentioned polynomial behavior near $t=0$ takes over and the
solution can be followed all the way to and across the
singularity. Some time after $t=0$, the story repeats and we can use
the adiabatic description again, leading to a decomposition of higher
oscillation state, which can be described in terms of Bogoliubov
transformations, as we will show in the following section. 

%% In other
%% words, if we only focus on each mode $A_n$, then we can think of the
%% singularity as a black box which transforms the amplitude of the
%% incoming solution $A_n(t\rightarrow-\infty)$ to a different amplitude
%% for $A_n(t\rightarrow+\infty)$. 

%% In terms of the $A_n$ coefficients, the fact that $\Psi(t)$ is an
%% straight line, translates into a relationship between the $in$ and
%% $out$ states, namely $\sum_n a_n^{in}=\sum_n a_n^{out}$ and $\sum_n
%% b_n^{in}=\sum_n b_n^{out}$. 

\begin{figure}[t!]
\begin{center}
\resizebox*{4.8in}{3in}
{\includegraphics{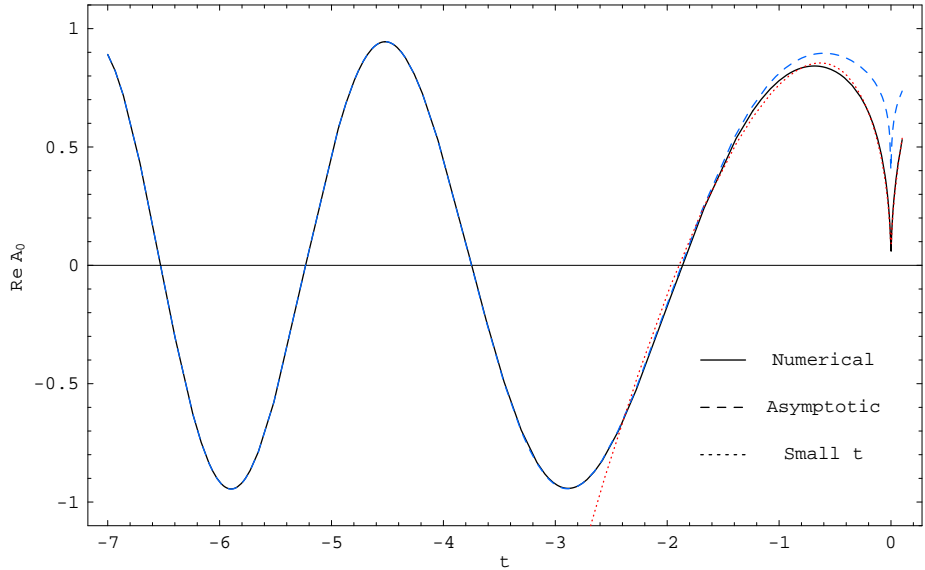}}
\caption{Using the orthogonality property (\ref{ortho}) one can
  extract the coefficients of the harmonic oscillator
  modes $A_n$ from the numerical solution of the wave function (see
  equation (\ref{harmonicbasis})). Here,
  we plot $Re(A_0)$ for $p_\phi=1$ and
  $p=0$. The dashed line represents the analytic solution
  (\ref{A0sol}) which follows the numerical solution (solid line) up
  to the place where the $t=0$ behavior, given by solution
  (\ref{psi-straight}) becomes more accurate (dotted line). We
  normalize the wave function so $A_0\sim 1$ initially. }
\lab{An2}
\end{center}
\end{figure}

\begin{figure}[t!]
\begin{center}
\resizebox*{4.8in}{3in}
{\includegraphics{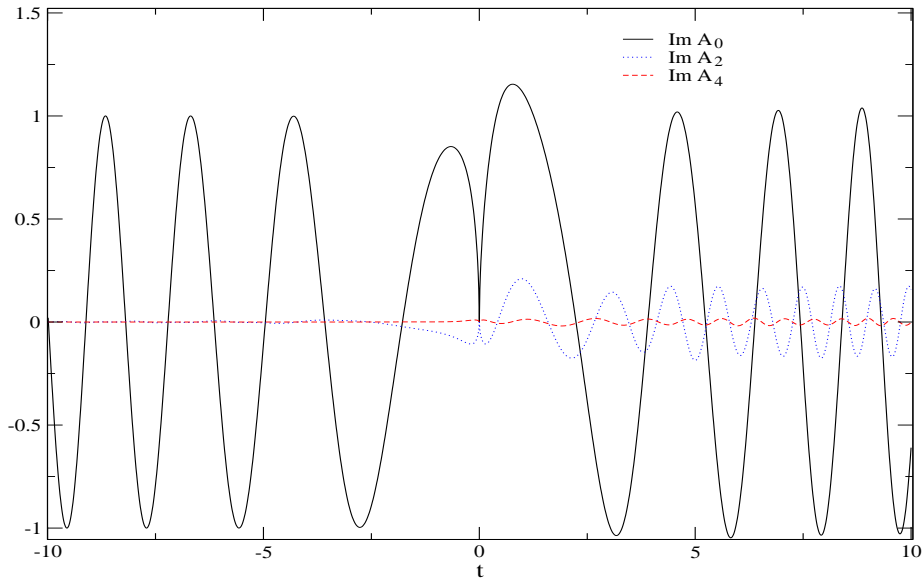}}
\caption{Using the orthogonality property (\ref{ortho}) we plot the $A_n$'s
  for different mode numbers $n$.  One can see how higher modes get
  excited after $t=0$, and the amplitude of higher states decrease with
 $n$ as expected. We use $p_\phi=0.793521$ and $p=0.4$. } 
\lab{An}
\end{center}
\end{figure}

\subsubsection{Particle production}\lab{partprodsection}

One has to be careful when constructing the Hilbert space, since 
a $C^{\infty}$ and globally hyperbolic manifold is needed. In
our case the string metric is not well defined at $t=0$, therefore we
can split the space of solutions into two sections: for $t<0$ and
$t>0$. Based on the regularity of the classical and quantum solutions
around $t=0$ it is natural to assume unitarity at $t=0$, which allows
us to use the Bogoliubov transformations to relate the two sections
of the Hilbert space and calculate particle production. Our numerical results will 
act to justify this assumption. Moreover, a natural inner product in this Hilbert space is
defined by \cite{birrell}
\be
(\Psi_1,\Psi_2)\equiv -i\int_\infty^\infty
dR\left[\Psi_1\frac{\der\Psi_2^*}{\der t}
  -\left(\frac{\der\Psi_1}{\der t}\right)\Psi_2^*\right].
\ee

One can expand the wave function $\Psi(R,t)$ in terms of two basis sets, one
for asymptotically negative times and one for the corresponding
positive times, namely
\ba
\Psi(R,t)&=&\sum_n a_n^{(in)}U_n^{(in)}+ (a_n^{(in)})^\dagger
(U_n^{(in)})^* \qquad \qquad t\rightarrow - \infty \nonumber \\ &=&
\sum_n a_n^{(out)}U_n^{(out)}+ (a_n^{(out)})^\dagger (U_n^{(out)})^* \qquad
\qquad t\rightarrow + \infty 
\ea
where $a_n^\dagger$ and $a_n$ are the creation and annihilation
operators associated with the harmonic oscillator expansion, and which obey
the usual commutation relationships (\ie $[a_n,a_m^\dagger]=\delta_{mn}$,
etc.). We would expect the negative time vacuum
($a_n^{(in)}\ket{ 0}_{in}=0$) to be different from that of positive
times ($a_n^{(out)}\ket{0}_{out}=0$), leading to particle
production. The functions $U_n^{(in)}$ and $U_n^{(out)}$ are the
positive frequency modes
\ba
U_n^{(in)}&=&\mathcal{N}_n \exp(+iE_n t) \exp(-|t|^sR^2/2)
H_n(|t|^{s/2}R) ~~~t<0,\nonumber \\
U_n^{(out)}&=&\mathcal{N}_n \exp(-iE_n t) \exp(-|t|^sR^2/2)
H_n(|t|^{s/2}R)~~~t>0,
\ea
with the normalization factor $\mathcal{N}_n$ given by 
\be\lab{normalization}
\mathcal{N}_n=(2^{n+1}n!\sqrt{(2n+1)\pi})^{-1/2},
\ee
in such a way that $(\Psi_n,\Psi_m)=\delta_{nm}$. 

Actually, since we are interested in having the vacuum as our
incoming state, the wave function for $t\rightarrow-\infty$ takes the form
\be\lab{incomingsol}
\Psi=\mathcal{N}_0\left[\left(a_0^{(in)}+(a_0^{(in)})^\dagger\right)\cos(E_0
|t|)-i\left(a_0^{(in)}-(a_0^{(in)})^\dagger\right)\sin(E_0
|t|)\right]\exp(-|t|^{s} R^2/2),
\ee
where $\mathcal{N}_0=\frac{1}{(4\pi)^{1/4}}$.
Now, we can write explicitly the wave function for
$t\rightarrow+\infty$ as
\ba
\Psi&=&\sum_{n} \mathcal{N}_n\bigg[\left(a_n^{(out)}+(a_n^{(out)})^\dagger\right)\cos(E_n
t)-i\left(a_n^{(out)}-(a_n^{(out)})^\dagger\right)\sin(E_n
t)\bigg]H_n(t^{s/2}R) e^{-t^{s} R^2/2} \nonumber
 \\ \nonumber &=& \sum_{n} \mathcal{N}_0
 \bigg[\left(a_0^{(in)}+(a_0^{(in)})^\dagger\right)\mathcal{D}_n \cos(E_0
   |t|+\varphi_n)\\ && \hspace{1.5cm} -i\left(a_0^{(in)}-(a_0^{(in)})^\dagger\right) 
   \bar{\mathcal{D}}_n \sin(E_0 |t|+\bar{\varphi}_n)\bigg]H_n(t^{s/2}R) e^{-t^{s} R^2/2},
\ea
where $\mathcal{D}_n $ and $\varphi_n$ ($\bar{\mathcal{D}}_n $ and
$\bar{\varphi}_n$) are the amplitude and phase --- with respect to
$t=0$ --- of the outgoing mode $n$ after sending an incoming {\it
  cosine} ({\it sine}) piece of the incoming wave function
(\ref{incomingsol}). Since the two lines in the previous
equation are equal, the coefficients should be related in the
following way 
\be
a_n^{(out)}=\alpha^*_{n0}a_0^{(in)}+\beta^*_{n0}(a_0^{(in)})^\dagger,
\ee
where the Bogoliubov coefficients are defined as
\be\lab{bogo}
\alpha_{n0}=\frac{\mathcal{N}_0}{2\mathcal{N}_n}\left(\mathcal{D}_ne^{i\varphi_n}
+\bar{\mathcal{D}}_ne^{i\bar{\varphi}_n}\right),\qquad\qquad
\beta_{n0}=\frac{\mathcal{N}_0}{2\mathcal{N}_n}\left(\mathcal{D}_ne^{i\varphi_n}
-\bar{\mathcal{D}}_ne^{i\bar{\varphi}_n}\right).
\ee
Particle production at a given energy level can then be understood in
terms of the expectation value of the particle number operator
$\hat{N}_n=(a_n^{(in)})^\dagger a_n^{(in)}$ over the out-vacuum
$\ket{0}_{out}$, which simplifies to the following
\be
\bra{0}\hat{N}_n\ket{0}_{out}=\beta_{n0}\beta_{n0}^*=2^{n-2}n!\sqrt{2n+1}
\left(\mathcal{D}_n^2+\bar{\mathcal{D}}^2_n-
\mathcal{D}_n\bar{\mathcal{D}}_n \cos(\varphi_n-\bar{\varphi_n})\right).
\ee
Figure \ref{partno} shows the particle production for different Kasner
exponents. A fit of this plot shows that the particle
production decays exponentially with mode number, $n$, satisfying a fit of 
the form $\beta_{00}^2 \exp(-C_1 n^{3/4})$, where $\beta_{00}$
and $C_1$ depend on the particular choice of Kasner exponents. Of
particular note is that for large $n$ where the energy per mode, $E_n$
satisfies  $E^2_n\sim n$,  
(see equation (\ref{En})), we then obtain an exponential decay of the
form $\beta_{00}^2 \exp(-C_1 E_n^{3/2})$, a result that agrees with
the semiclassical instanton calculation in \cite{Mturok} 
for the Milne universe, but remains to be compared for other Kasner 
exponents.

\begin{figure}[t!]
\begin{center}
\resizebox*{4.8in}{3in}
{\includegraphics{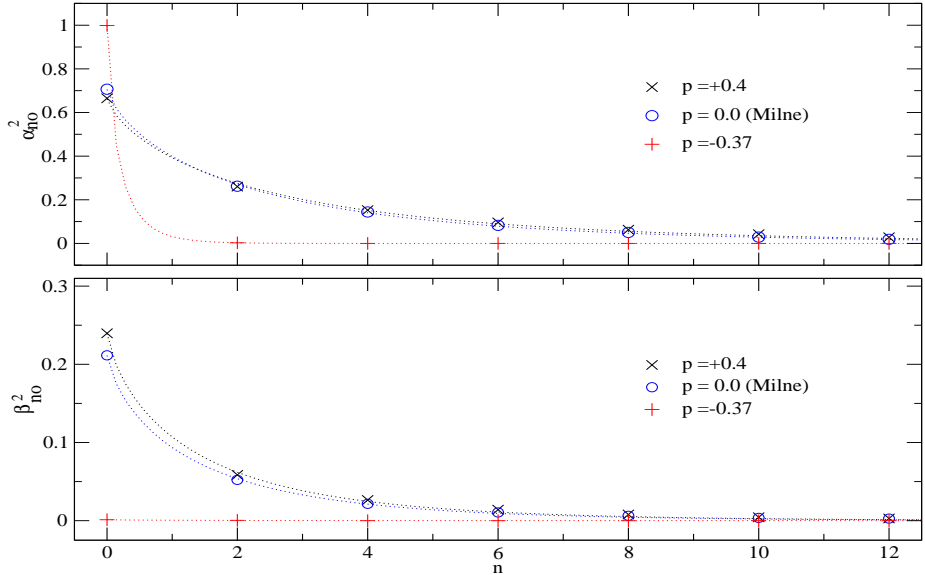}}
\caption{Bogoliubov coefficients $\alpha_{n0}^2$ and $\beta_{n0}^2$  as functions of the mode number $n$ for three
  Kasner exponent combinations: $p=0.4$, $p=0.0$ (Milne) and $p=-0.37$, with
  $p_\phi$ obtained using (\ref{pphisol}) for $m=7$. In the last case,
the effective (3+1)-metric where the string oscillates is almost flat,
thus the particle production $|\beta|^2$ is negligible. A fit shows
that both Bogoliubov coefficients decay as $C_1
\exp (-C_2 n^{3/4})$, where $C_1$ and $C_2$ are positive constants
that depend on the Kasner exponents only. } 
\lab{partno}
\end{center}
\end{figure}

\subsection{Non-zero center of mass momentum}\lab{nozerop}

If one starts with a non-vanishing center of mass momentum, $\pi_z$, at
$t=-t_0$, and if the orbifold rapidity $\theta_0$ is not negligible with respect
to $\pi_z$, then one cannot drop its contribution in the
Hamiltonian (\ref{hamcom}). The Wheeler-de Witt equation associated with the Hamiltonian
(\ref{hamcom}) can be solved numerically, again yielding a regular solution, 
in particular at $t=0$, which is very similar to the $\pi_c=0$ case
shown in Fig.~\ref{wavefn}. Furthermore, to describe the $t>0$
behavior, we can once again use the harmonic oscillator basis as an
ansatz, but with a frequency which now depends on both time and
$\pi_c$; details are given in Appendix A. Here we only summarize
our findings. Asymptotically, the wave function can   
again be described in terms of decoupled harmonic oscillator modes, while 
near the singularity the straight-line behavior (after integrating
over $R$) remains as a characteristic feature. Thus, we can send in the
vacuum state and read off the harmonic oscillator components in the
outgoing solution, to find the Bogoliubov coefficients for such
a transition. 

From the definition in equation (\ref{pi_c}), we know that the orbifold rapidity
should play a role in the physical observables, via the combination
${\pi}_o=\theta_0^{(2\Delta p +s)/(s+2)}\pi_z$. Thus, if as mentioned
above we assume $\Delta p>-s/2$, increasing the orbifold rapidity is
equivalent to increasing the center of mass momentum. Now the question
is: how do the Bogoliubov coefficients change due to $\pi_c$? On the
one hand, we know 
that classically the presence of a singularity forces the loop to
increase its velocity until it reaches the speed of light at $t=0$. As
we have already seen, this speeding up is achieved  
by exciting higher order oscillation modes. Therefore, one would
naively expect that a larger $\pi_c$ 
should help the loop to reach the speed of light at $t=0$ without
exciting higher order modes. On the other hand, one should expect that
increasing the orbifold rapidity makes the orbifold collision more
violent, thus increasing particle production. These two effects
compete against each other and indeed, we see both of them at
different scales. For small $\pi_c$, particle production is 
suppressed exponentially as $\exp(-c_1\pi_c^2-c_2 n^{3/4})$, where 
$c_1>0$ and $c_2>0$ depend on the Kasner exponents, and $c_2$ has a mild
dependence on $\pi_c$. However, for larger $\pi_c$ particle production
grows as a powerlaw of $\pi_c$, while 
exponentially decreases with $n$. This powerlaw growth is in
agreement with the instanton calculation of \cite{Mturok}, where it was
found that $\bra{0}\hat{N}_0\ket{0}_{out}\propto \theta_0^{(d-1)/3}$ for a
$(d+1)$-dimensional spacetime. In agreement, we find for the case $d=4$; three
spatial dimensions corresponding to the effective space in which the
loop oscillates plus the M-theory dimension. Figure 
Fig.~\ref{partnopz} shows how particle production decays with $n$, and
Fig.~\ref{partnopz2} depicts how the particle number of the ground
level, $\bra{0}\hat{N}_0\ket{0}_{out}$, changes with $\pi_c$. At
first sight, it may be worrying to think that the particle production
diverges for large $\pi_c$, but actually one should have in mind that
the orbifold rapidity should be small, in order to have modes created
well inside the comoving Hubble horizon where the adiabatic regime can
be trusted \cite{Mturok,niz}. Furthermore, the transverse velocity of any 
loop cannot be the speed of light, otherwise it could not oscillate in
the $xy$ plane. 

For a gas of loops, the average velocity should be close to
$1/\sqrt{2}$, as shown in \cite{turokvel} for flat space\footnote{We
  can assume a flat background initially, since for a small $\theta_0$
  the size of the loops is small compare to the Hubble radius .}. We
can then imagine a single loop with such a velocity at $t=-t_s$ (see
equation (\ref{t0})), and calculate the corresponding value of $\pi_c$
using the first equation in (\ref{eqnscirc}) and
(\ref{pi_c}). Assuming the loop is initially static
($\dot{R}(-t_s)=0$) and of unit size (${R}(-t_s)=1$), one obtains 
\be\lab{vcm}
\pi_c=\frac{v\theta_0^{(2\Delta p +2)/(s+2)}}{\sqrt{1-v^2}}.
\ee
If $\theta_0$ is small and the velocity dispersion around the mean
value is not large, then the particle production should be small. For
example, for $\theta_0\sim 1/5$ and a velocity dispersion of around $\pm 0.25$
(\cite{press}), the corresponding values of $\pi_c$ lie on the range
$\epsilon\in (0.3,1.93)$ (at $t=-t_s=-5$), which maps to the exponentially
decaying region of Fig.~\ref{partnopz2}.

\begin{figure}[hpt!]
\begin{center}
\resizebox*{4.8in}{3in}
{\includegraphics{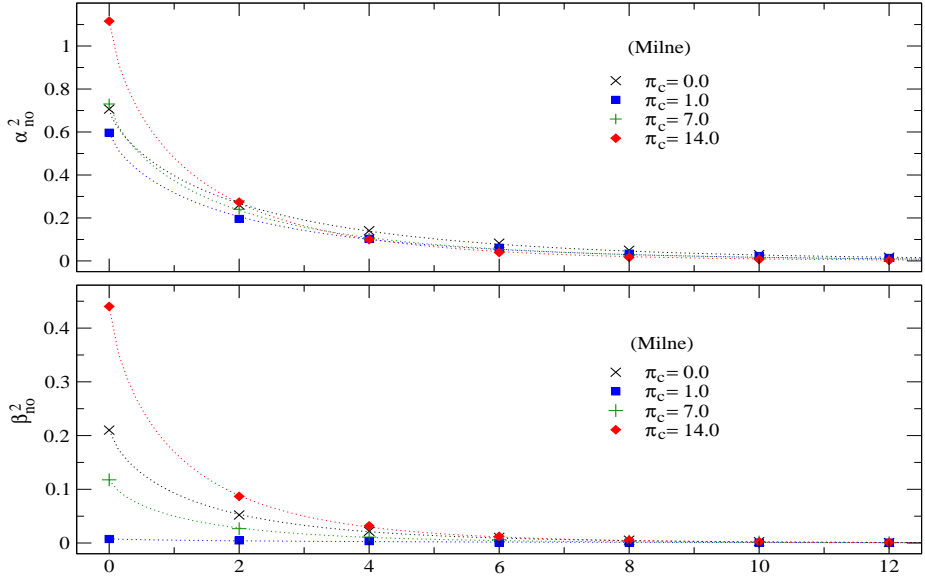}}
\caption{Bogoliubov coefficients $\alpha_{n0}^2$ and $\beta_{n0}^2$
  as functions of the mode number $n$ for four different initial
  momenta, $\pi_z$, measured at $t=-t_0=-10$. The Milne
  universe ($p=0$, $p_\phi=1$) is assumed for all cases. Fits show that both
  coefficients decay as $c_2\exp(-c_1 n^{3/4})$, where $c_1$ has a
  mild dependence on $\pi_c$, but $c_2$ strongly depends on $\pi_c$
  as shown in Figure \ref{partnopz2}.}
\lab{partnopz}
\end{center}
\end{figure}

\begin{figure}[hpt!]
\begin{center}
\resizebox*{3.6in}{2.25in}
{\includegraphics{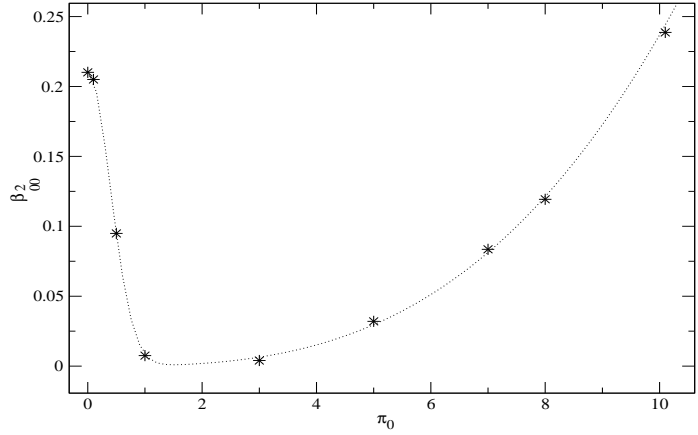}}
\caption{Bogoliubov coefficient $\beta_{00}^2$ for different initial
  initial momentum $\pi_c$ chosen at $t_s=5$, and assuming the Milne
  universe ($p=0$, $p_\phi=1$). For small center of mass
  momentum the loop needs less excited modes to reach the speed of
  light at $t=0$, but a large $\pi_c$ is equivalent to a large
  orbifold rapidity, which increases the particle production due the
 more violent brane-collision. The fitting curve has a
  form $c_2\exp (-c_1 \pi_c^{2})+c_3 \pi_c^3$, where the $c_i$'s are
  positive constants.} 
\lab{partnopz2}
\end{center}
\end{figure}

Finally, an important check of the quantum model is to verify
unitarity, which implies that the canonical commutation relationships should be
preserved over time. This translates into the following property
for the Bogoliubov coefficients 
\be
1=\sum_n  \alpha_{n0}\alpha_{n0}^*-\beta_{n0}\beta_{n0}^* .
\ee
One would need to sum over all states to get unity, however, one can
see that using the first ten excited modes the number is close to
unity (in Figure (\ref{qprod})).

\begin{figure}[hpt!]
\begin{center}
\resizebox*{4.8in}{3in}
{\includegraphics{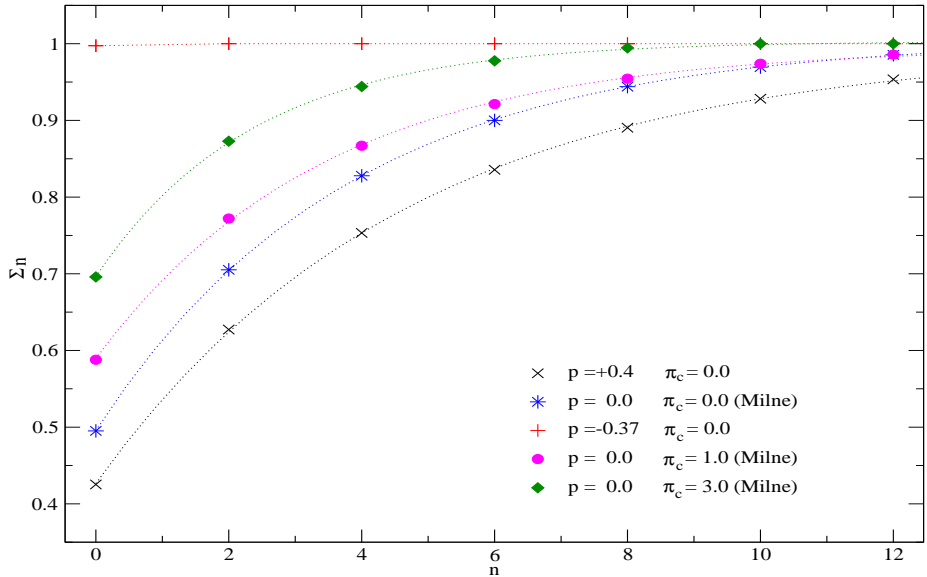}}
\caption{Plot of $\sum_n=\sum_{i=0}^n (\alpha_{i0}^2-\beta_{i0}^2)$
  as a function of the mode
  number $n$, for the same three Kasner metrics of Fig.~\ref{partno},
  and for two non-zero values of $\pi_c$ at $t_s=5$. Unitarity is
  preserved if $\sum_n\rightarrow 1$ as $n\rightarrow\infty$.} 
\lab{qprod}
\end{center}
\end{figure}

\section{Conclusions} \lab{sec:discuss}

The nature of the big bang is one of the biggest problems facing
cosmology. No one fully understands the physics that occurred in this
crucial period. One of the more interesting recent proposals for the
origin of the big bang, is through the collision and re-emergence of
two orbifold planes in eleven dimensions \cite{ekp, cyc, Mturok}. As
the two planes approach each other, the light states of the theory
consist of winding M2-branes, which are described in terms of
fundamental strings in a ten dimensional background. Near the brane 
collision region, the full eleven-dimensional metric considered is that of Euclidean space times a
compactified 1+1-dimensional Milne universe. In \cite{niz}, two of us
considered the the classical evolution of winding membranes in such a
background, showing that they  suffered no blue-shift as the M theory
dimension collapses, and their equations of motion remained regular
across the transition from big crunch to big bang. However, one may
expect there to really be small 
perturbations to the background metric, leading to a more general
Kasner background for the light membranes to evolve in. If this is the
case, an obvious question is what happens to collapsing light
M2-branes as they pass through the singularity?  In this 
paper we have gone beyond to original  classical analysis of winding membranes in
\cite{niz} to include general Kasner backgrounds. By considering  the corresponding Hamiltonian
equations, we have been able to solve for the wave function of loops with circular
symmetry and demonstrate the sensitivity of the solutions to the values of the Kasner
exponents $p$ around the Milne solution ($p=0$ and
$p_\phi=1$). As is evident from Figures \ref{plots} and
\ref{moreplots}, although the general behavior remains similar, in
particular the loop solutions remain perfectly finite, there is clear
evidence that the amplitude and periodicity of the outgoing modes
depend strongly on the precise value of the Kasner exponents. In the
regime $p\sim-p_\phi/2$, the energy density has a mild dependence on
$|t|$,  and the modes do not feel the contraction or expansion of the
universe, leading to no classical energy production.  On the other
hand, for large positive values of $p$, the string does feels the
effects of the evolving  scale factor, which produces a large effect
around $t=0$, and a greater classical production of energy. To confirm
the particle interpretation of this classical result, we adopt the
Wheeler-de Witt formalism to quantize the system of the evolving
circular loop in the Kasner background. Circular symmetry of the loop
is crucial in order to help us solve the system of equations as it
allows us to assume $\partial_\sigma t =0$, implying the time
coordinate of the loop is a function of $\tau$ only. Remarkably we are
able to solve the Wheeler-de Witt equation both far from and close to
the singularity $t=0$. Asymptotically, the loop is well described
through a WKB approximation, given by
Eqns.~(\ref{harmonicbasis}-\ref{vacuumsol}) in the case of very small
orbifold rapidity compared to the center of mass momentum, or by
Eqns.~(\ref{A-eqn-app})-(\ref{en_po}) for the general case. 
Of particular note though is the fact that near the singularity, the
solution simplifies to such an extent that the integrated wave
function (\ref{intewave}) is given by a straight line in $t$ through
the singularity which allows a 
simple understanding of the solution on either side of $t=0$, made
evident through Figures \ref{wavefn} and \ref{An2}. The
complicated non-periodic evolution seen in Figure \ref{wavefn} just
after the singularity provides evidence that there could well be
particle production as a loop evolves through a singularity. This is
confirmed in our analysis of the particle production, as seen in
Figures \ref{partno} and \ref{partnopz} where we plot the particle
production number for three separate values of the Kasner parameter
including the usual Milne case, and three different center of mass
momenta. In all cases, particle production is exponentially suppressed
with the oscillation mode number $n$, and in the case of zero-center
of mass momentum $\pi_z$ it is independent of the orbifold rapidity
$\theta_0$. However, for a non-vanishing center of mass momentum,
particle production depends on the simple function
$\pi_c=\pi_c(\theta_0,\pi_z)$, defined in (\ref{pi_c}). Production of
particles is exponentially suppressed for small $\pi_c$, but has 
a powerlaw growth for large $\pi_c$. The two effects are expected:
the first one corresponds to the fact that a small initial
velocity helps the loop to reach the speed of light at the singularity
without exciting higher order modes. In contrast, the second effect
corresponds to the fact that a large orbifold rapidity induces a more 
violent big crunch/big bang, which results in more particles been
produced.  

As we have seen, as the loop passes through the singular point, higher
order modes become light and excited, the string reaches the speed of
light everywhere along it, and the effect is either an increase or
decrease in the amplitude of the outgoing mode leading to classical gain or
loss in energy and the production of particles whilst maintaining
Unitarity.  

There are of course caveats to what we have done in this paper. In
reality the Kasner metric is not flat away from the singularity and
thus it can receive corrections when interacting with the membrane. It
is not clear how this would affect the background and subsequent
analysis? Assuming we are only slightly away from Milne case, we
believe that the analysis should be similar to what we have done here.   

It is very encouraging that we have seen how it is possible to have
finite particle production in through a singular region in a Kasner
background. It is now worth seeing quite how much reheating occurs in
such a scenario. An obvious, if somewhat difficult calculation is to
extend this work beyond the case of a simple circular loop. Finally,
it will be interesting to see how these results can be interpreted
from the conformal theory description of \cite{hertog}.

{\bf Acknowledgements.} We would like to thank Jorma Louko, Stephen
Creagh, Sven Gnutzmann, Karima Righetti and Gregor Tanner 
for useful conversations, and also Alkistis Pourtsidou for 
reading the manuscript. EJC is grateful to the Royal Society, 
and GN is grateful to STFC  for financial support.  

\appendix

\section{Non-vanishing $\pi_c$ solution} \lab{sec:appendixA} 

To solve the Wheeler-de Witt equation derived from the Hamiltonian (\ref{hamcom}) for a non-vanishing $\pi_c$ term, one can use the same ideas as in the $\pi_c=0$
case. First and to get a felling of the solution, one could study the wave equation derived from (\ref{hamcom}) for a constant time $t=t_0$, namely 
\be
\hat{H}\Psi=\left(\partial_t^2-
\partial_R^2+\tilde{t}_0^{2\Delta p}\pi_c^2+\tilde{t}_0^{2s}R^2\right)\Psi=0,  
\ee
where $\pi_c$, $\Delta p$ and $s$ are defined in (\ref{pi_c}) and (\ref{sdef}). A solution to 
this equation is given again by the harmonic oscillator solution
(\ref{asym-wave-func}), but with an energy
$E_n^2=(2n+1)t_0^{s}+\pi_c^2t_0^{\Delta p}$. Therefore, we can use
the same ansatz (\ref{harmonicbasis}) and the properties of the
Hermite polynomials to get a recursive differential equation for the
function $A_n(t)$, which is similar to (\ref{A-eqn}) but with extra terms
containing $\pi_c$.  Actually, in the $T$-time variable, the $A_n$
equation reads
\be\lab{A-eqn-app}
0=\frac{d^2{A}_n}{dT^2}+(1+2n)A_n+\pi_c^2\left(\frac{s+2}{2}|T|^{\frac{2}{s+2}}\right)^{2\Delta
  p-s}A_n+\mathcal{O}\left(\frac{1}{T}\right), 
\ee
which does not have a simple analytical solution for a generic Kasner
exponent combination, unlike the $\pi_c=0$
case. In the particular case of the Milne universe, the solution to
this equation is given in terms of Whittaker functions, which are
related to the polylogarithm functions. The appearance of these functions
suggests a consistent description with the $1/\alpha'$-series solution
of classical evolution near $t=0$, where polylogarithm functions were
found to second order in the string tension \cite{niz}.

Since we are searching the asymptotic spectrum of states, it is enough
to find a series solution of equation (\ref{A-eqn-app}). The expansion
parameter is $Q\equiv\pi_c^2\left(\frac{s+2}{2}|T|^{\frac{2}{s+2}}\right)^{2\Delta
  p-s}=\pi_c^2 |t|^{2\Delta p-s}$, and the solution becomes more
accurate for large $|T|$ because (\ref{regcond}) implies $-2\Delta
p+s=p_\phi+2p_z\geq 0$. This series solution can be constructed using
the ansatz 
\be
A_n=\exp(i\mathcal{E}_nT),
\ee
 where
\be\lab{en_po}
\mathcal{E}_n\equiv \mathcal{E}_n(T)=\sqrt{\beta+\alpha Q
 +\gamma Q^2}.
\ee
The following constants 
\be
\beta=2n+1,\qquad \alpha=\frac{s+2}{2-s+2\Delta p}, \qquad \gamma= 2\frac{s-2\Delta p}{2-s+2\Delta
p},
\ee
solve equation (\ref{A-eqn-app}), with an error of order
$T^{-(4\Delta p+s+2)/(s+2)}\sim t^{-(4\Delta p+s+2)/2}$. So, we can use this approximate solution and 
calculate the particle production using the same expressions as in
section \ref{partprodsection}. We assume the two vacua are given by
\ba
\Psi(R,t)&=&\sum_n a_n^{(in)}U_n^{(in)}+ (a_n^{(in)})^\dagger
(U_n^{(in)})^* \qquad \qquad t\rightarrow - \infty \nonumber \\ &=&
\sum_n a_n^{(out)}U_n^{(out)}+ (a_n^{(out)})^\dagger (U_n^{(out)})^* \qquad
\qquad t\rightarrow + \infty 
\ea
where 
\ba
U_n^{(in)}&=&\mathcal{N}_n \exp(+i{E}_n t) \exp(-|t|^sR^2/2)
H_n(|t|^{s/2}R) ~~~t<0,\nonumber \\
U_n^{(out)}&=&\mathcal{N}_n \exp(-i{E}_n t) \exp(-|t|^sR^2/2)
H_n(|t|^{s/2}R)~~~t>0,
\ea
and $E_n=\frac{2}{s+2}\mathcal{E}_n|t|^{s/2}$. The normalization factor $\mathcal{N}_n$ is given by 
\be
\mathcal{N}_n=\left[2^{n+1}n!\sqrt{\pi}\left(\mathcal{E}'_nT 
+\mathcal{E}_n\right)\right]^{-1/2},
\ee
with $\mathcal{E}'_n\equiv \frac{d}{dT}\mathcal{E}_n(T)$. Therefore, by writing the vacuum solution as
\be\lab{incomingsol-app}
\Psi=\mathcal{N}_0\left[\left(a_0^{(in)}+(a_0^{(in)})^\dagger\right)\cos({E}_0
|t|)-i\left(a_0^{(in)}-(a_0^{(in)})^\dagger\right)\sin({E}_0
|t|)\right]\exp(-|t|^{s} R^2/2),
\ee
we can read off the outgoing solution in terms of the Bogoliubov
coefficients (\ref{bogo}), and calculate the particle production in the same way
we did for the $\pi_c=0$ case. The results are shown in Figure
\ref{partnopz} and summarized in Section \ref{nozerop}.

\end{document}